\documentclass[3p,12pt,times]{elsarticle}
\usepackage{amssymb}
\usepackage{amsthm}
\usepackage{amsmath}
\usepackage[OT2,T1]{fontenc}
\DeclareSymbolFont{cyrletters}{OT2}{wncyr}{m}{n}
\DeclareMathSymbol{\Sha}{\mathalpha}{cyrletters}{"58}
\usepackage[figuresright]{rotating}
\usepackage{subfig}

\usepackage{graphicx}
\usepackage{dcolumn}
\usepackage{bm}
\usepackage[colorlinks = true,
linkcolor = blue,
urlcolor  = blue,
citecolor = blue,
anchorcolor = blue]{hyperref}
\usepackage{url}
\usepackage{lineno}
\biboptions{sort&compress}

\usepackage{changes}

\begin{document}
	\begin{frontmatter}
		\title{Analysis of multistability in discrete quantum droplets and bubbles}
		
		\author[au1,au1b]{R.\ Kusdiantara}
		\author[au5]{H.\ Susanto}
		\ead{hadi.susanto@yandex.com}
		\author[au5]{T.F.\ Adriano}
		\author[au4]{N.\ Karjanto\corref{cor1}}
		\ead{natanael@skku.edu}
		
		\address[au1]{Industrial and Financial Mathematics Research Group, Institut Teknologi Bandung,\\ Jl.\ Ganesha No.\ 10, Bandung, 40132, Indonesia}
		\address[au1b]{Centre of Mathematical Modelling and Simulation, Institut Teknologi Bandung,\\ Jl.\ Ganesha No.\ 10, Bandung, 40132, Indonesia}
		\address[au5]{Department of Mathematics, Khalifa University, PO Box 127788, Abu Dhabi, United Arab Emirates}
		\address[au4]{Department of Mathematics, University College, Natural Science Campus, Sungkyunkwan University,\\ 2066 Seobu-ro, Jangan-gu, Suwon, Gyeonggi-do, 16419, Republic of Korea}
		\cortext[cor1]{Corresponding author}

		\begin{abstract}
				This study investigates the existence and stability of localized states in the discrete nonlinear Schr\"odinger (DNLS) equation with quadratic and cubic nonlinearities, describing the so-called quantum droplets and bubbles. Those states exist within an interval known as the pinning region, as we vary a control parameter. Within the interval, multistable states are connected through multiple hysteresis, called homoclinic snaking. In particular, we explore its mechanism and consider two limiting cases of coupling strength: weak (anti-continuum) and strong (continuum) limits. We employ an asymptotic and a variational method for the weak and strong coupling limits, respectively, to capture the pinning region's width. The width exhibits an algebraic and an exponentially small dependence on the coupling constant for the weak and strong coupling, respectively.  This finding is supported by both analytical and numerical results, which show excellent agreement. 
				We also consider the modulational instability of spatially uniform solutions. Our work sheds light on the intricate interplay between multistability and homoclinic snaking in discrete quantum systems, paving the way for further exploration of complex nonlinear phenomena in this context. 
		\end{abstract}
		
		
	\end{frontmatter}
	
	
	\section{Introduction}
	
	While one-dimensional solitons of the nonlinear Schr\"odinger (NLS) equation are robust localized wave packets and the ground state of the model~\cite{novikov1984theory}, the corresponding multidimensional solitons, including fundamental and higher-order states with topological structures, are different. In this multidimensional setting, the common cubic self-attractive nonlinearity typically yields completely unstable solitons~\cite{fibich2015nonlinear,kartashov2019frontiers}. This instability arises from the tendency for critical and supercritical collapse in the respective dimensions, leading to singular solutions via catastrophic wave collapse (self-compression)~\cite{sulem2007nonlinear}. Therefore, a significant focus lies on developing physically relevant scenarios enabling stabilization of both fundamental and topologically structured multidimensional self-trapped localized modes as they would offer characteristics that would be inaccessible in the one-dimensional case, such as higher-dimensional self-trapped states with topological charge (intrinsic vorticity)~\cite{malomed2019vortex}.
	
	A promising approach to address this challenge was recently proposed theoretically in~\cite{petrov2015quantum}, based on utilizing quantum fluctuations as a correction to the mean-field behavior of Bose-Einstein condensates (BEC), which was initially forecasted by Lee, Huang, and Yang~\cite{lee1957eigenvalues}. The proposal was subsequently realized experimentally in quantum droplets, magnetic quantum gases, and dipolar supersolids; see, e.g., the reviews \cite{chomaz2022dipolar,bottcher2020new}. The Lee-Huang-Yang effect manifests through local quartic self-repulsive terms within the corresponding NLS equations in the mean-field approximation. A localized state that would typically collapse is now sustained by this interplay between mean field attraction and Lee-Huang-Yang repulsion, and one has a superfluid state characterized by a density constrained to remain below a specific maximum value, rendering it incompressible. Hence, this quantum macroscopic state is classified as a fluid and is referred to as ``quantum droplets''~\cite{luo2021new}.
	
	In cigar-shaped and pancake-shaped BECs, which appear in a three-dimensional setting, each system can be effectively described by one-dimensional and two-dimensional NLS equations, respectively. Consequently, under tight confinement imposed by an external potential in one direction or more, one can change the form of the effective NLS equation and create lower-dimensional droplets. In the one-dimensional limit, the Lee-Huang-Yang term exhibits quadratic self-attraction instead of being quartic in the original three-dimensional setup~\cite{petrov2016ultradilute}. 
	
	Dynamics and characteristics of one-dimensional quantum droplets are studied in~\cite{astrakharchik2018dynamics}, whereby upon solving the associated amended Gross-Pitaevskii equation, the authors identified two physically distinct regimes that correspond to small droplets of a Gaussian-like shape and large puddles with a broad flat-top plateau. Furthermore, Zhou et al.\ \cite{zhou2019dynamics} studied them in the presence of linear optical lattice, where they interestingly observed multi-stability and stability change between onsite and off-site solitons. It was later shown that the multi-stable droplets are connected in a multi-hysteresis fashion~\cite{dong2020multi} and exist in a finite interval of band-gap. Kartashov and Zezyulin \cite{kartashov2024enhanced} also considered bright solitons in a linear optical lattice and showed enhanced mobility with alternating mobility and immobility bands related to soliton multistability. Additionally, Zhao et al.\ \cite{zhao2021discrete} considered when the lattice is deep, and the NLS equation effectively becomes a discrete model, demonstrating that multistability is also present. 
	
	The multistability reported in \cite{zhou2019dynamics,dong2020multi,zhao2021discrete} belongs to the phenomenon of pinning observed in many nonlinear dynamical systems with bistability of two or more uniform states (see, e.g., \cite{susanto2011variational,matthews2011variational} and references therein). In such systems, a front linking these states typically undergoes directional drift, depending on which state is preferred. However, there is no preference between the two states at a specific parameter value, resulting in a stationary front. This point is known as the \emph{Maxwell point}, occurring when the two states have equal energy. Placing two fronts back-to-back creates a localized state, and the bifurcation diagram, plotting the length of this localized solution against a control parameter, exhibits a snaking structure. This structure involves a sequence of turning point bifurcations in near-perfect alignment over a range of parameter values around the Maxwell point, forming what is known as the pinning region. Since the spatial structure of such a localized state departs from and returns to a uniform state, this phenomenon is named \emph{homoclinic snaking}~\cite{woods1999heteroclinic}. 
	
	The present paper aims to analyze homoclinic snaking and pinning regions of discrete quantum droplets. We also consider localized solutions called holes/dark notches/quantum bubbles. The main results of our report are the mechanism behind the snaking and the asymptotic width of the pinning region of the two solution kinds in two limiting cases, i.e., strong and weak coupling between the wave packets. 
	
	The paper is structured as follows. We introduce the model for the dynamics of discrete quantum droplets in Section~\ref{sec2}. We also analyze the uniform solutions of the model in the same section. We analyze the existence of localized states, i.e., discrete quantum droplets and quantum bubbles, in Section~\ref{sec3}. The multistability of these states that form homoclinic snaking is also reported numerically in the section. We present our asymptotic analysis of the snaking in Section~\ref{sec4}. Finally, we conclude our work in Section~\ref{sec5}.
	
	\section{Mathematical model and uniform solutions}	           \label{sec2}
	
	In this study, we consider the following discrete NLS equation with quadratic and cubic nonlinearity, i.e.,
	\begin{equation}
		i\partial_t \phi_n=-\mu \phi_n-\frac{C}{2}\left(\phi_{n+1}-2\phi_n+\phi_{n-1}\right)-|\phi_n|\phi_n+ |\phi_n|^2\phi_n,		\label{eq:dnls23}
	\end{equation}%
	where $\phi_n(t)$ represents the complex-valued wave packet amplitude of the quantum droplet, $C$ represents coupling strength between wave packets, and $\mu$ denotes the chemical potential that, in the following discussion, will be used as a bifurcation parameter. 
	\textcolor{black}{Both $C$ and $\mu$ are dimensionless parameters in this context, allowing us to generalize our results to various physical systems by scaling appropriately based on specific experimental or theoretical needs.}
	Equation~\eqref{eq:dnls23} has potential (Hamiltonian) as follows:
	\begin{equation}
		E(\phi)=\sum_{n=-\infty}^{\infty} \left[\frac{\mu}{2}|\phi_n|^2+\frac{C}{2}\left|\phi_n-\phi_{n-1}\right|^2+\frac{1}{3}|\phi_n|^3-\frac{1}{4}|\phi_n|^4\right].             \label{eq:E_dnls23}
	\end{equation}
	Next, consider the time-independent (real-valued) solution of Equation\ \eqref{eq:dnls23}, i.e., 
	\begin{equation}
		0=-\mu \phi_n-\frac{C}{2}\left(\phi_{n+1}-2\phi_n+\phi_{n-1}\right)-|\phi_n|\phi_n+ \phi_n^3,		\label{eq:dnls23_ti}
	\end{equation}
	which is also known as the {discrete Allen-Cahn equation}. Here, the double vertical lines denote the absolute value instead of the modulus from the complex-valued case.
	
	Generally, once a real-valued solution $\phi_n=\tilde{u}_n$ is obtained, its linear stability can be determined by perturbing it as $\phi_n=\tilde{u}_n +\epsilon (\Bar{v}_n(t)+i\Bar{w}_n(t))$. 
	After substituting it 
	into \eqref{eq:dnls23}, linearizing around $\epsilon=0$, and separating the real and imaginary parts of the resulting equations, we then take $\Bar{v}_n(t) = \hat{v}_n e^{\lambda t}$ and $\Bar{w}_n(t) = \hat{w}_n e^{\lambda t}$, from which we obtain the following linear eigenvalue problem:
	\begin{equation}
		\lambda \left(\begin{array}{c}
			\hat{v}_n\\\hat{w}_n
		\end{array}\right) = \underbrace{\left(\begin{array}{cccc}
				0 & -\mathcal{L}_{-} \\
				\mathcal{L}_{+} & 0 \\
			\end{array}\right)}_{\mathcal{L}} 
		\left(\begin{array}{c}
			\hat{v}_n\\\hat{w}_n
		\end{array}\right), \label{eig}
	\end{equation}
	where the anti-diagonal matrix entries are given as follows:
	\begin{equation}
		\begin{array}{rcl}
			\mathcal{L}_{-} &=& \mu + \dfrac{1}{2}C\Delta +  |\tilde{u}_n| -  \tilde{u}_n^2, \vspace*{0.25cm}\\
			\mathcal{L}_{+} &=&  \mu + \dfrac{1}{2}C\Delta + 2|\tilde{u}_n| - 3\tilde{u}_n^2.
		\end{array}
	\end{equation}
	A solution is said to be linearly stable when $\text{Re}\left(\lambda \right)\leq0$ for all the eigenvalues and unstable when at least one of the spectra has $\text{Re}\left(\lambda \right)>0$.
	
	\begin{figure}[tbhp!] 
		\centering
		\includegraphics[scale=0.55]{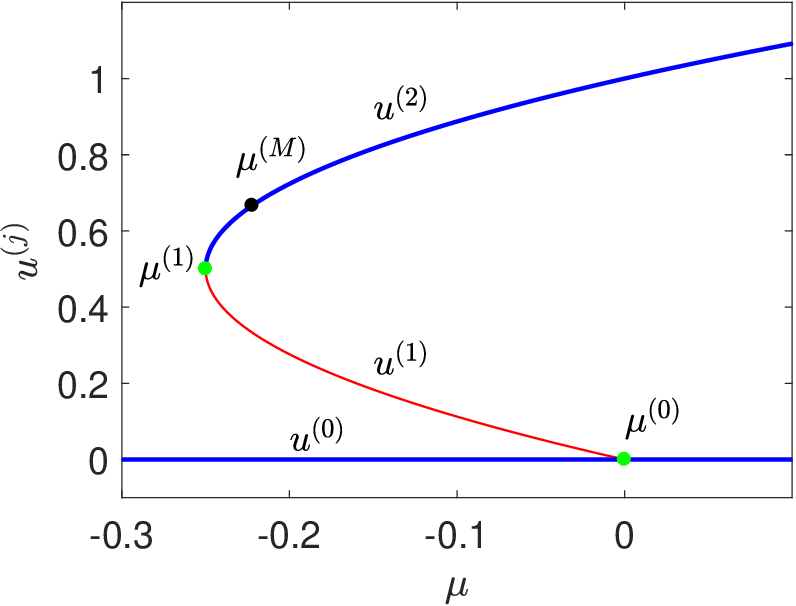}
		\caption{
			\textcolor{black}{(Color online) Uniform solution of Equation\ \eqref{eq:dnls23} as a function of the chemical potential $\mu$. For point forward, in the bifurcation diagram, the thick blue and thin red curves indicate the stable and unstable solutions, respectively. The green points indicate bifurcation points, i.e., pitchfork bifurcation $\mu^{(0)}$ and saddle-node bifurcation $\mu^{(1)}$. The critical value $\mu^{(M)}=-2/9$ represents the Maxwell point, where the energies of competing stable states are equal, thus defining a threshold separating different dynamical regimes within the diagram.}
		}
		\label{fig:unisol}
	\end{figure}
	
	Equation \eqref{eq:dnls23} has uniform solutions (homogeneous states) $\phi_n(t)=u^{(j)}$, $j=0,1,2$, given by
	\begin{equation}
		u^{(0)}=0,\quad u^{(1)}=\frac{1-\sqrt{4\mu+1}}{2},\quad \text{and}\quad u^{(2)} = \frac{1+\sqrt{4\mu+1}}{2}.    \label{eq:unisol}
	\end{equation}
	The spatially uniform solution $u^{(1)}$ bifurcates from $u^{(0)}$ at the parameter value $\mu=\mu^{(0)}=0$ and collides with the solution $u^{(2)}$ at $\mu=\mu^{(1)}=-1/4$. 
	We plot the bifurcation diagram in Figure \ref{fig:unisol}.
	
	\textcolor{black}{
		This figure illustrates the uniform solutions of Equation~\eqref{eq:dnls23}  as they vary with the chemical potential $\mu$. The diagram is color-coded to enhance clarity and interpretation. The thick blue curves represent the stable solutions to the equation, where the system tends to remain steady and exhibits predictable behavior over time. Conversely, the thin red curves denote the unstable solutions. In these regions, small perturbations in $\mu$ can lead to significant changes in the system's behavior, indicating sensitivity to initial conditions and parameter values. \vspace{0.25cm}\\
		The green points marked on the diagram are critical as they signify the locations of bifurcations, which are fundamental changes in the structure of the solution space as a function of the parameter $\mu$. Specifically, the green point labeled $\mu^{(0)}$ represents a pitchfork bifurcation, where a single solution branches into multiple solutions, i.e., $u^{(0)}$ and $u^{(1)}$, typically indicating a symmetry-breaking in the system as $\mu$ varies. \vspace{0.25cm}\\ 
		The other green point, labeled $\mu^{(1)}$, corresponds to a saddle-center bifurcation, where two solution branches, i.e., $u^{(1)}$ and $u^{(2)}$, converge and annihilate each other, leading to a sudden disappearance of the solution types as $\mu$ crosses this value.	\vspace{0.25cm}\\	
		Overall, the bifurcation diagram provides critical insights into the dynamics governed by Equation~\eqref{eq:dnls23}, illustrating how solutions evolve and change stability as the chemical potential $\mu$ is adjusted. Such diagrams are essential for predicting the system's behavior under various conditions and understanding the theoretical underpinnings of the modeled phenomena.
	}
	
	To determine the linear stability of the uniform solution $u^{(j)}$, we substitute $\hat{v}_n,\,\hat{w}_n\sim e^{ikn}$, where $k$ is the wave number of the perturbation, into the eigenvalue problem \eqref{eig} from which we obtain the dispersion relation
	\begin{equation}
		\lambda(k)=\pm \left[\left(- \mu - C\left(\cos(k)  - 1\right)   + \left|u^{(j)}\right|- \left({u^{(j)}}\right)^2 \right)\left(\mu+ C\left(\cos(k)  - 1\right) - 2\left|u^{(j)}\right| + 3\left({u^{(j)}}\right)^2\right)\right]^{\frac{1}{2}}.		\label{eq:disper}
	\end{equation}
	
	\begin{figure}[tbhp!] 
		\centering
		\subfloat[]{\includegraphics[scale=0.55]{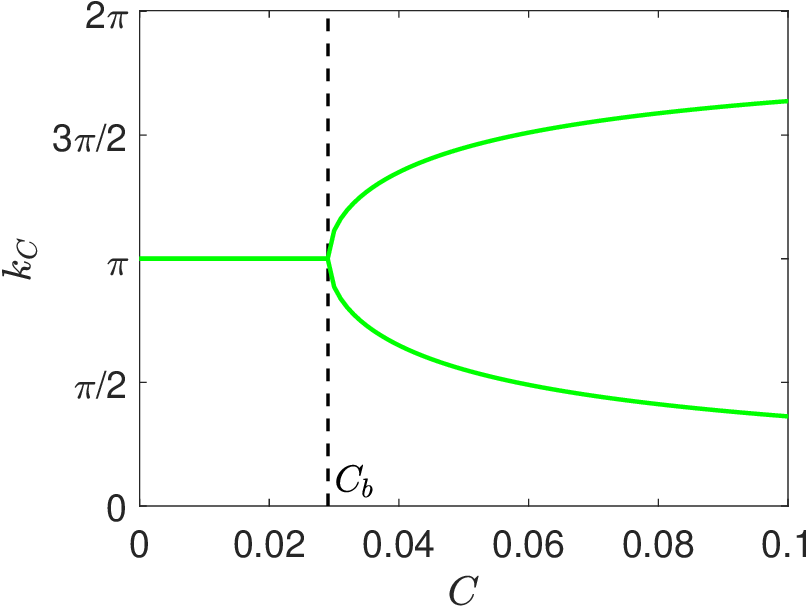}\label{subfig:k_crit_mu_0_15}}\quad
		\subfloat[]{\includegraphics[scale=0.55]{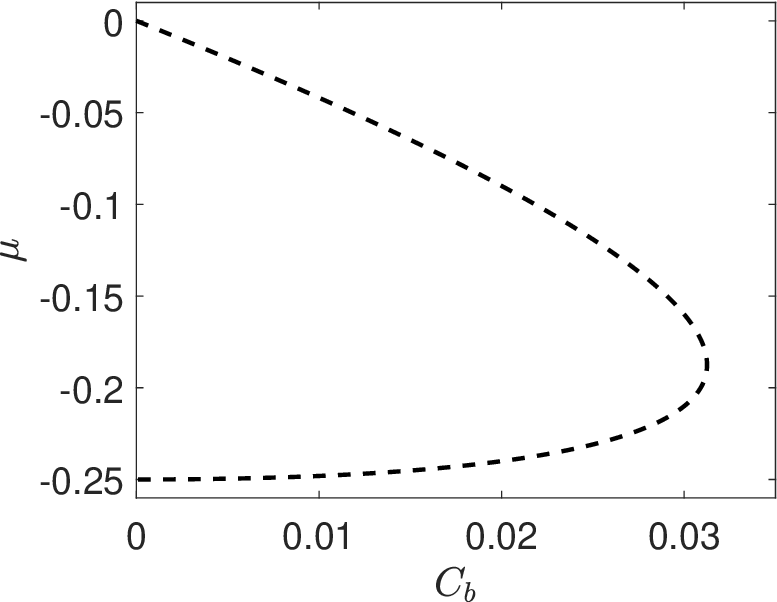}\label{subfig:mu_vs_Cb}}
		\caption{\textcolor{black}{
				(a) Depiction of the critical wave number, $k_C$, as defined in Equation~\eqref{eq:kc}, plotted as a function of the parameter $C$ for $\mu = -0.15$ and $j=1$. The result identifies a specific critical value, $C_b = 0.0291$.
				(b) The critical value $C_b$, as derived from Equation~\eqref{eq:cb}, plotted as a function of $\mu$.	
		}}          
		\label{fig:k_crit}
	\end{figure}
	
	A uniform solution is said to be stable when $\lambda(k)\leq 0$ for $\forall \; k \in \mathbb{R}$ and unstable when $\exists \; k$ such that $\lambda(k)>0$. The maximum of the spectrum \eqref{eq:disper} is at
	\begin{equation}
		k_C=\pm\left\{\begin{array}{ccc}
			\pi+2m\pi&,&C\leq C_b\\
			\arccos\left(\dfrac{-2\mu+2C-3{u^{(j)}}+4\left({u^{(j)}}\right)^2}{2C}\right)+2m\pi&,&C>C_b
		\end{array}\right.,\quad m \in \mathbb{Z},
		\label{eq:kc}
	\end{equation}
	where
	\begin{equation}
		C_b=\frac{1}{2}\mu+\frac{3}{4}{u^{(j)}}-\left({u^{(j)}}\right)^2.
		\label{eq:cb}
	\end{equation}
	We plot this critical wavenumber \textcolor{black}{$k_C$} in Figure~\ref{subfig:k_crit_mu_0_15}. The figure illustrates the relationship between the critical wave number, $k_C$, as defined in Eq.\ \eqref{eq:kc}, and the parameter $C$. 
	At the threshold $C_b$, there is a splitting where afterward the dispersion relation \eqref{eq:disper} has two maxima.
	\textcolor{black}{
		For $\mu = -0.15$, we observe that 
		$C_b = 0.0291$. Figure\ \ref{subfig:mu_vs_Cb} plots the values of $C_b$ within the existence interval of the unstable uniform solution $u^{(1)}$, i.e., $\mu^{(1)} < \mu < \mu^{(0)}$. 
	}
	
	
	\begin{figure}[tbhp!] 
		\centering
		\subfloat[$C=0.01$]{\includegraphics[scale=0.55]{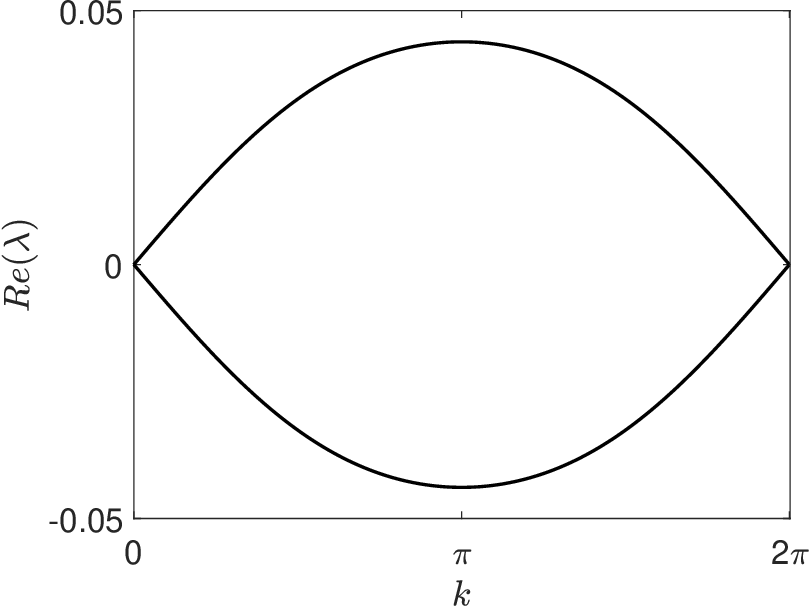}}
		\subfloat[$C=0.05$]{\includegraphics[scale=0.55]{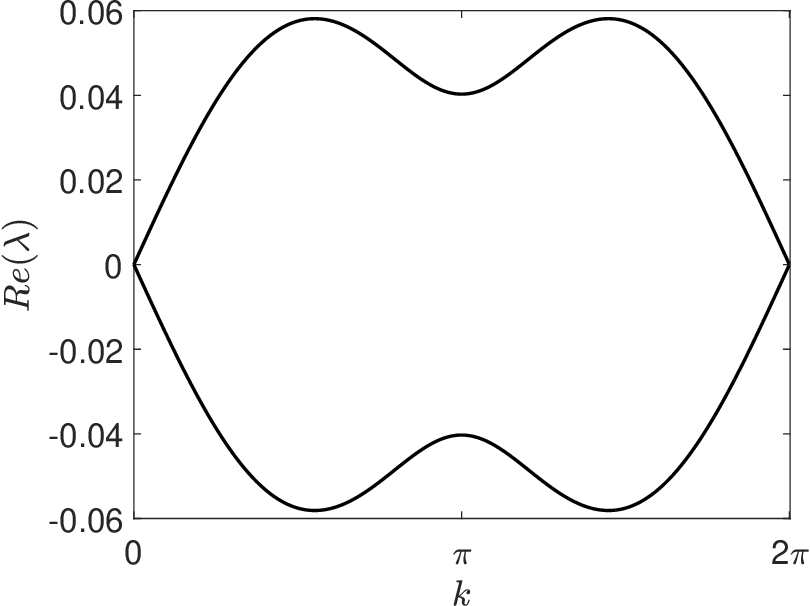}}
		\caption{Dispersion relation of lower uniform solution $u^{(1)}$ for $\mu=-0.15$. The real part of $\lambda$ vs.\ the wavenumber $k$ is shown. The left and right panels correspond to $C = 0.01$ and $C = 0.05$, respectively.
		}  
		\label{fig:disrel}
	\end{figure} 
	
	\begin{figure}[tbhp!] 
		\centering
		\subfloat[$C=0.01$]{\includegraphics[scale=0.55]{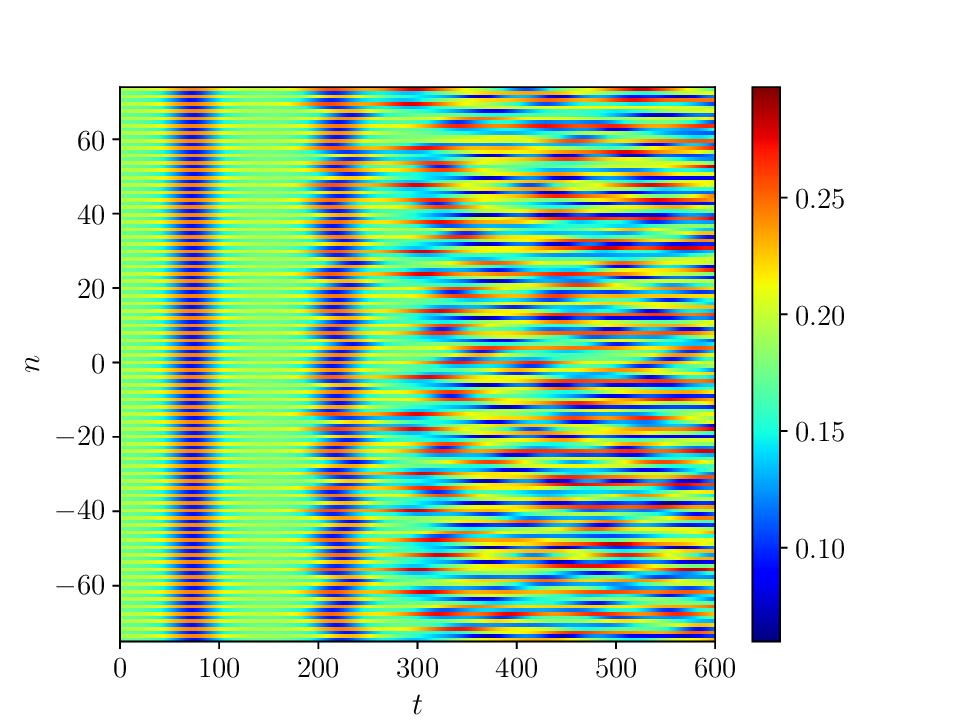}}
		\subfloat[$C=0.05$]{\includegraphics[scale=0.55]{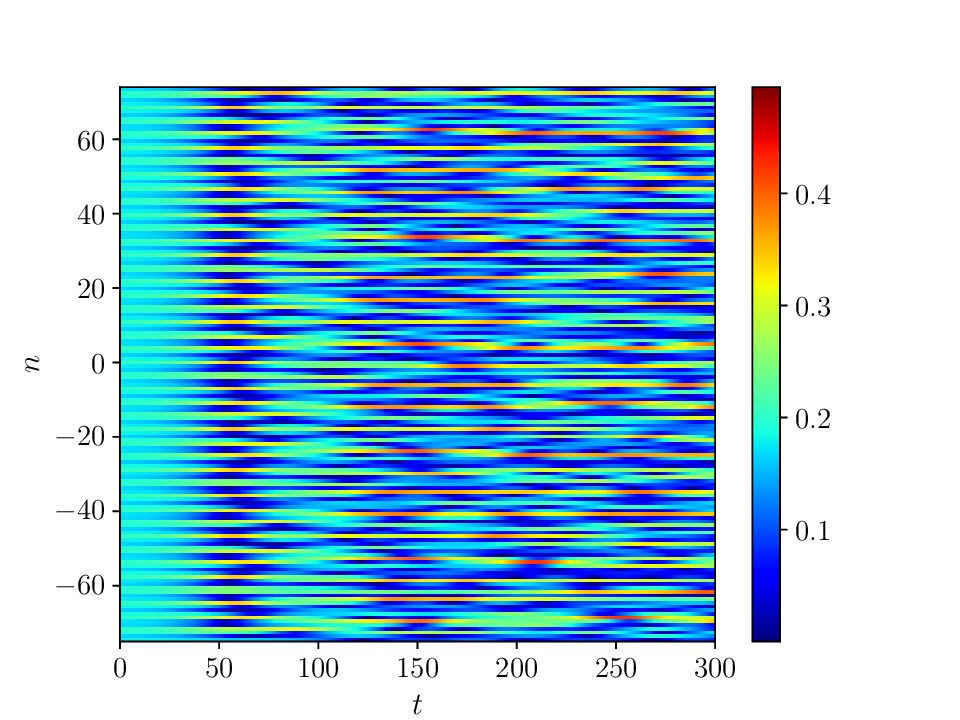}}
		\caption{Top view of the typical time dynamics of the unstable uniform solution $u^{(1)}$ for $\mu=-0.15$ and $C=0.01$ and $0.05$, \textcolor{black}{after small random perturbations initially}.}   \label{fig:dynamic_kc}
	\end{figure}

	We obtain that the zero solution $u^{(0)}$ is always stable because $\lambda$ is purely imaginary, i.e.,
	\begin{equation}
		\lambda(k) = \pm i\left|\mu +C\left(\cos(k)-1\right)\right|.
	\end{equation}
	The upper state $u^{(2)}$ is also stable in its entire region. Therefore, we have a bistability of uniform solutions for $\mu^{(1)}\leq \mu$. Their potential is given by
	\begin{equation*}
		E\left({u^{(j)}}\right)=\frac{\mu}{2}\left({u^{(j)}}\right)^2-\frac{1}{4}\left({u^{(j)}}\right)^4+\frac{1}{3}\left({u^{(j)}}\right)^3.    
	\end{equation*}
	For the two uniform solutions, 
	the potential vanishes simultaneously at $\mu=\mu^{(M)}=-\dfrac{2}{9}$, a critical parameter called the {Maxwell point}. 
	For $\mu < \mu^{(M)}$, $E\left({u^{(0)}}\right)<E\left({u^{\textcolor{black}{(2)}}}\right)$ and hence, the zero state is more favorable than the ``upper'' state, and vice versa. 
	
	For the uniform solution $u^{(1)}$, which exists for $\mu^{(1)}\leq \mu \leq \mu^{(0)}$, we plot its dispersion relation in Figure \ref{fig:disrel} for two different values of $C$ that correspond to the dispersion relation with one and two maxima. We obtain that $u^{(1)}$ is always unstable. 
	\textcolor{black}{Conversely, the uniform solution $u^{(2)}$ is always stable because the real part of the dispersion relation Equation~\eqref{eq:disper} is always zero.}
	We plot the typical time evolution of the unstable state in Figure \ref{fig:dynamic_kc} that corresponds to the dispersion relation in Figure \ref{fig:disrel}. Nonlinear dynamics of the modulational instability in the continuum limit $C\to\infty$ were discussed in~\cite{mithun2020modulational}. 
	
	\section{Localized solutions}       \label{sec3}
	
\begin{figure*}[tbhp!] 
		\centering
		\subfloat[$\mu=-0.23$]{\includegraphics[scale=0.37]{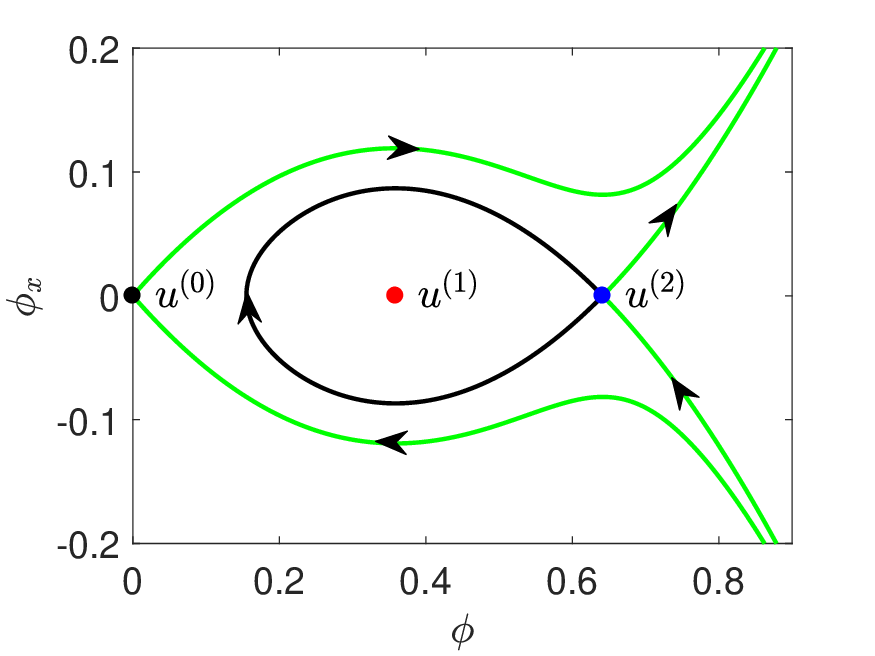}\label{subfig:phase_Mp_left}}
		\subfloat[$\mu={\mu^{(M)}}$]{\includegraphics[scale=0.37]{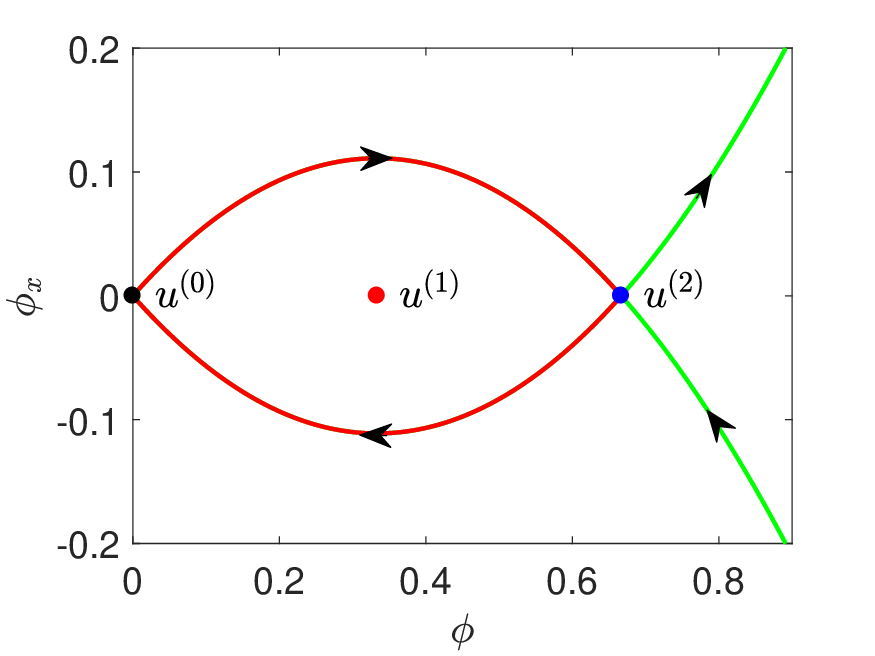}\label{subfig:phase_Mp}}
		\subfloat[$\mu=-0.21$]{\includegraphics[scale=0.37]{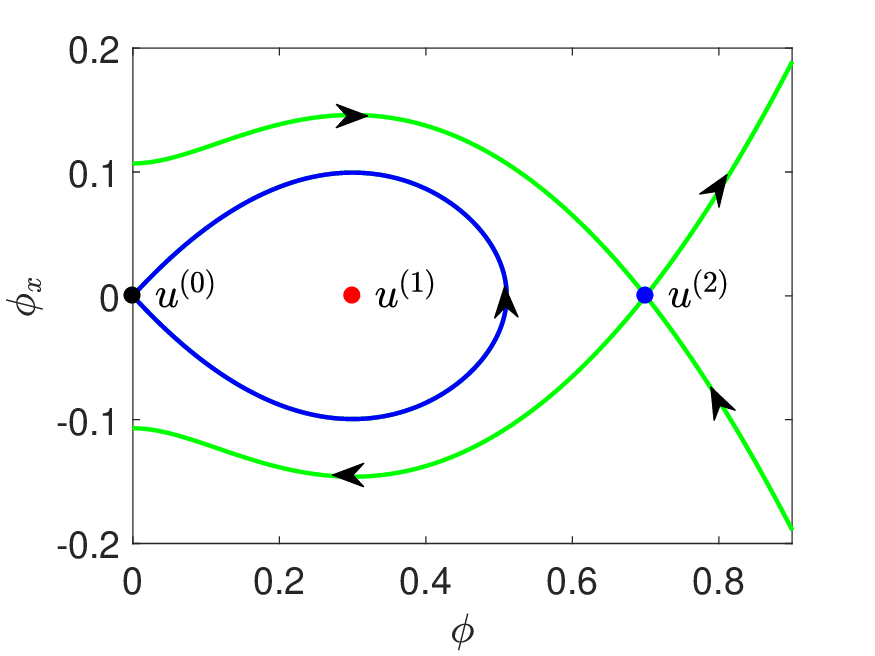}\label{subfig:phase_Mp_right}}
		\caption{Phase portrait of Equation~\eqref{eq:dnls23_ti} in the continuum limit ($C\rightarrow\infty$), i.e., Equation \eqref{eq:nls23_ti}, around the Maxwell point ${\mu^{(M)}}=-2/9$. The black, blue, and red curve trajectories represent the localized state's bright soliton, dark soliton, and kink. The green curves indicate other stable or unstable manifolds.}
		\label{fig:phase_potrait}
	\end{figure*}

	The discrete NLS equation~\eqref{eq:dnls23} admits localized solutions. \textcolor{black}{Depending on the setup of the localized solutions, we can have bright solitons (i.e., quantum droplets) or dark soliton solutions (i.e., holes or quantum bubbles) that are homoclinic to 
		the uniform solution ${u^{(0)}}$ and ${u^{(2)}}$, respectively. The bright and dark solitons exist in the same parameter value, i.e., \(\mu^{(1)} < \mu < \mu^{(0)}\)}.
	\textcolor{black}{To illustrate this, consider the continuum limit where \( \sqrt{C} \to \infty \) and \( \phi_n = \phi(x = n/\sqrt{C}) \). Under these conditions, Equation \eqref{eq:dnls23_ti} can be expressed as:
		\begin{equation}
			0 = -\mu \phi - \frac{1}{2} \phi_{xx} - |\phi|\phi + \phi^3. \label{eq:nls23_ti}
		\end{equation}
		Here, \( \phi_{xx} \) denotes the second spatial derivative of \( \phi \), reflecting the wave function's linearized curvature in the continuum limit, which plays a pivotal role in shaping the system's dynamics. We plot its phase portrait in Figure~\ref{fig:phase_potrait}. Homoclinic connections for \(\mu\) around the Maxwell point \(\mu^{(M)}\) encircle the uniform solution \({u^{(1)}}\) as the center.
	}

	
	For $\mu>{\mu^{(M)}}$, the homoclinic trajectory emanates from ${u^{(0)}}$, which indicates that it is a bright soliton, and bifurcates from ${\mu^{(0)}}$, see Figure\ \ref{subfig:phase_Mp_right}. For $\mu<{\mu^{(M)}}$, the emanation point of the trajectory is ${u^{(2)}}$, which corresponds to a dark soliton, and it bifurcates from ${\mu^{(1)}}$, see Figure\ \ref{subfig:phase_Mp_left}.	At the Maxwell point $\mu={\mu^{(M)}}$, we have a kink solution, which is represented by a heteroclinic trajectory, see Figure\ \ref{subfig:phase_Mp}. See also \cite{edmonds2023dark} for the transition from dark to bright solitons as one varies $\mu$. In the following, we will study the continuation of the solitons for finite $C$.
	
	\subsection{Bright solitons}

	
	
	First, we consider bright solitons of the discrete NLS equation \eqref{eq:dnls23} that bifurcate from the zero solution ${u^{(0)}}$ at point ${\mu^{(0)}}$. We are particularly interested in two fundamental localized solutions, i.e.,\ onsite and offsite states. In the uncoupled limit $C=0$, they are formed by the two bistable states, i.e., the non-zero solution ${u^{(2)}}$ as the ``upper'' state at several sites and the zero state ${u^{(0)}}$ as the ``background'' state. 
	For nonzero coupling $C$, we solve the time-independent Equation~\eqref{eq:dnls23_ti} using a Newton-Raphson method combined with a numerical continuation to vary $\mu$. At the same time, once a localized solution is obtained, we also solve the corresponding eigenvalue problem \eqref{eig} to determine its stability. 
	
\begin{figure}[tbhp] 
		\centering
		\subfloat[$C=0.1$]{\includegraphics[scale=0.5]{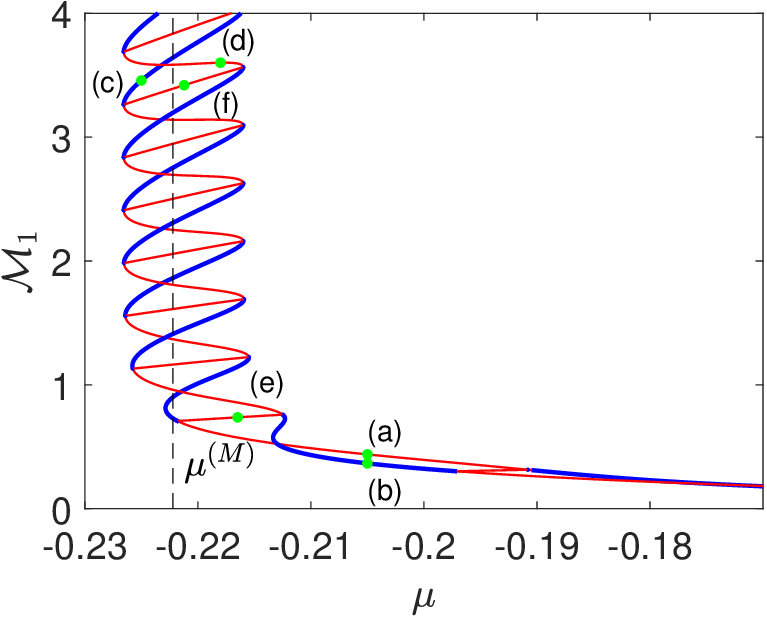}\label{subfig:bright_snake_c_0_10}} \qquad 
		\subfloat[$C=1$]{\includegraphics[scale=0.5]{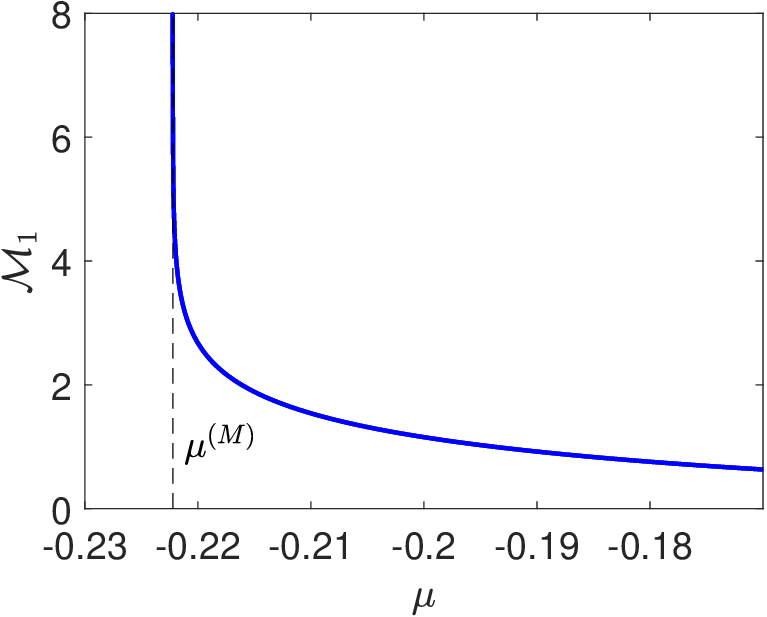}}
		\caption{
			\textcolor{black}{
				Bifurcation diagram of bright solitons for (a) \(C=0.1\) and (b) \(C = 1\). We plot the norm \(\mathcal{M}_1\) for varying \(\mu\). So-called ladders interconnect two snaking branches for onsite and offsite solitons. The thick blue lines represent stable solutions, while the thin red lines depict unstable solutions. The corresponding solutions at points depicted in panel (a) are plotted in Figure \ref{fig:bright_snakes_prof}. Additionally, the critical value \( \mu^{(M)} \) represents the Maxwell point, where the energies of competing stable states are equal, defining a threshold that demarcates different dynamical regimes within the diagram.
		}           	}
		
		\label{fig:bright_snakes}
	\end{figure}

	We plot our numerical results in Figure\ \ref{fig:bright_snakes}, that are depicted in terms of the squared ${l}^2$-norm
	\begin{equation}
		\mathcal{M}_1=\sum_{n}\left|\phi_n\right|^2,
	\end{equation}
	as a function of the parameter $\mu$. There are two main curves, which correspond to the (symmetric) onsite and offsite solutions. We obtain the multistability of localized states connected through turning point bifurcations, i.e., homoclinic snaking, within a finite interval of $\mu$, i.e., the pinning region. 
	
\begin{figure*}[tbhp] 
		\centering
		\subfloat[]{\includegraphics[scale=0.5]{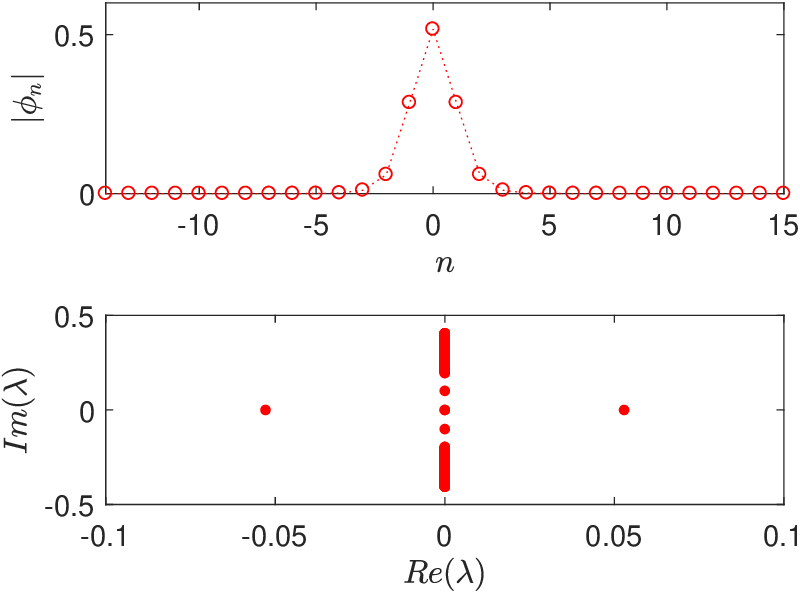}\label{subfig:prof_bright_a}}\,
		\subfloat[]{\includegraphics[scale=0.5]{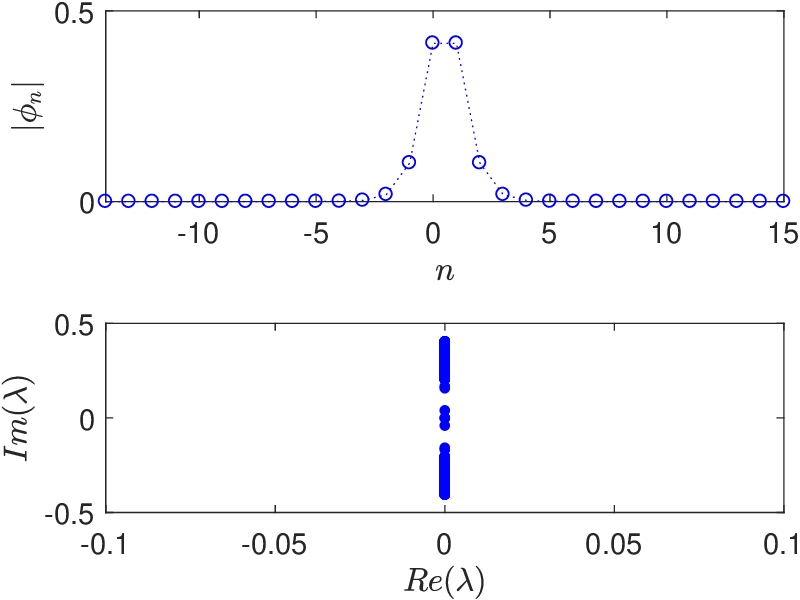}}\\
		\subfloat[]{\includegraphics[scale=0.5]{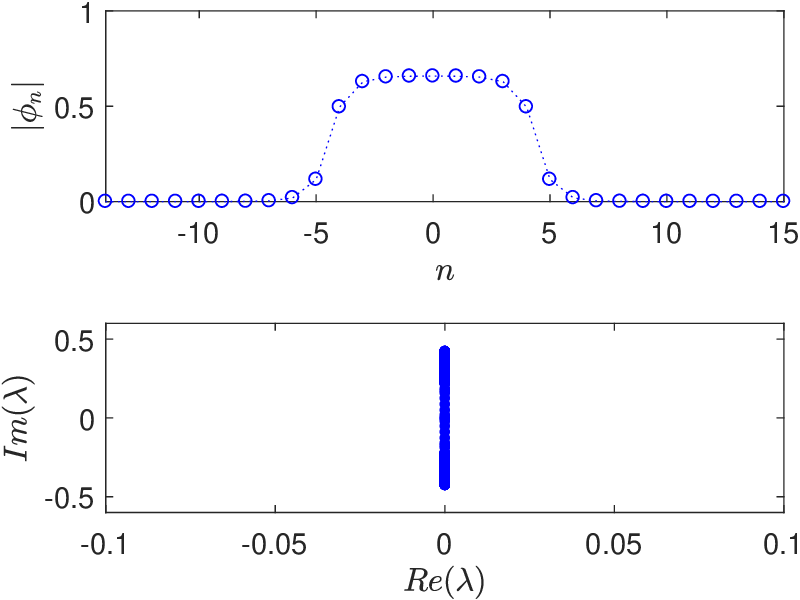}}\,
		\subfloat[]{\includegraphics[scale=0.5]{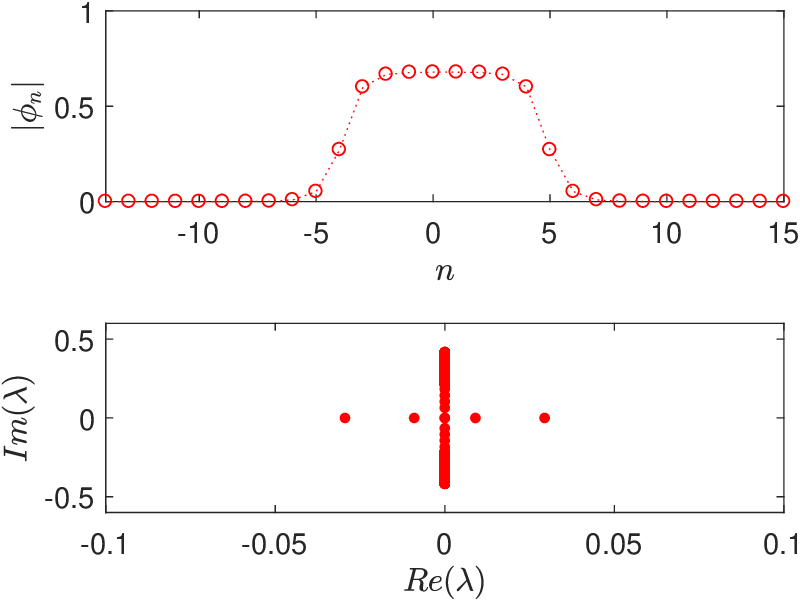}}\\
		\subfloat[]{\includegraphics[scale=0.5]{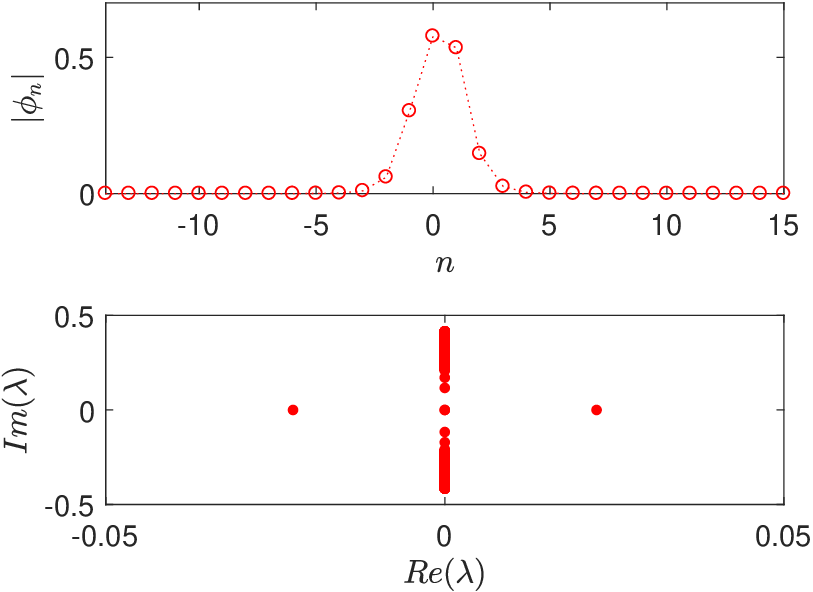}}\,
		\subfloat[]{\includegraphics[scale=0.5]{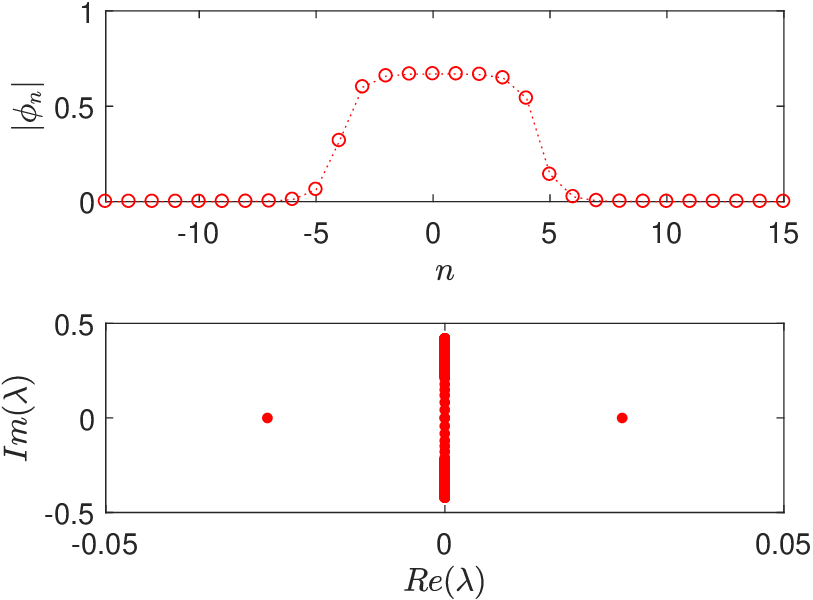}}
		\caption{(Color online) Plot of onsite and offsite solution profiles on the bifurcation diagram in Figure\ \ref{fig:bright_snakes} and their spectra in the complex plane. {Panels [(a), (c)],  [(b), (d)],  and [(e), (f)] are onsite, offsite, and ladder solutions, respectively}. Blue and red represent stable and unstable solutions, respectively, consistent with the color scheme used in Figure~\ref{fig:bright_snakes}.}                  		\label{fig:bright_snakes_prof}
	\end{figure*}

	\textcolor{black}{We also present in Figures \ref{fig:bright_snakes_prof} onsite, offsite, and asymmetric localized solutions along the diagram in Figure\ \ref{fig:bright_snakes} at several values of $\mu$. Onsite solitons, in the discrete setting, are localized modes centered on a lattice site, typically exhibiting maximum amplitude at a single lattice point, resulting in a symmetric profile around this site. On the other hand, offsite solitons are localized modes centered between two lattice sites, often showing less obvious symmetry than onsite solitons, with their maximum amplitude occurring between lattice points. Asymmetric states refer to more complex localized structures that do not exhibit symmetry and can span multiple lattice sites in irregular configurations. }
	
	As the bifurcation parameter $\mu$ varies,	the norm $\mathcal{M}_1$ increases accordingly, where the ``upper" state part of the localized solution invades the ``lower'' state background. Moreover, for a larger value of coupling strength $C$, the pinning region shrinks toward 
	the Maxwell point ${\mu^{(M)}}$. 
	In addition to symmetric solutions corresponding to the main branches, we also obtain asymmetric ones. In the bifurcation diagrams, they form ladders that connect the two main branches and are always unstable. Some are also plotted in Figure \ref{fig:bright_snakes_prof}.

	\begin{figure}[tbhp] 
		\centering
		\subfloat[]{\includegraphics[scale=0.5]{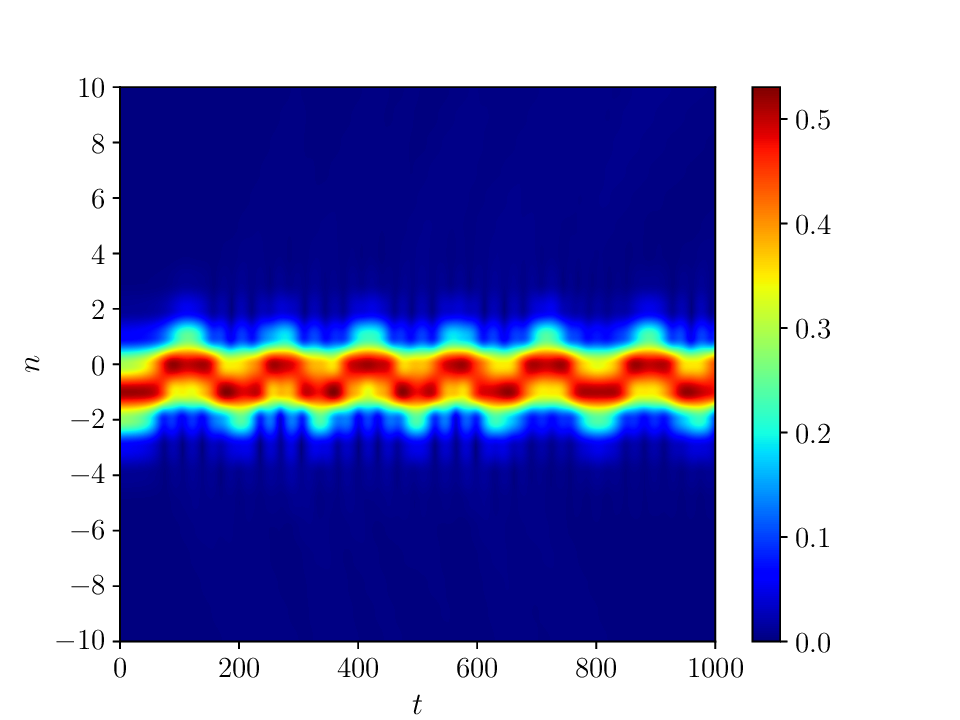}\label{fig:dynamics_bright_a}}\hspace{-1cm}
		\subfloat[]{\includegraphics[scale=0.5]{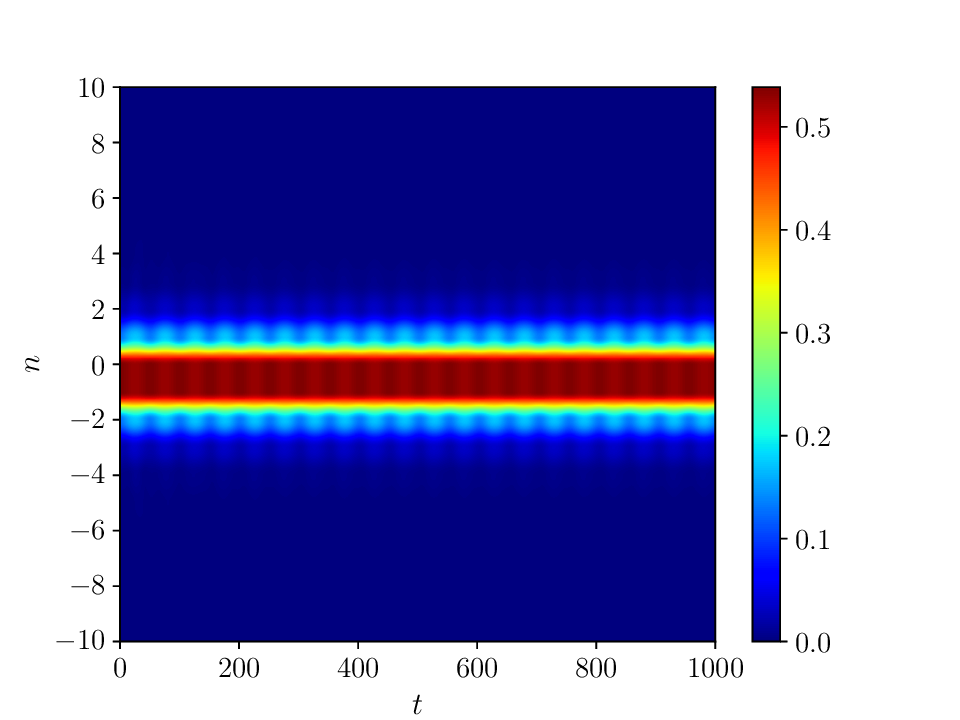}\label{fig:dynamics_bright_b}}
		\caption{Top view of the typical time dynamics of (a) unstable and (b) stable bright solitons. The initial profiles are taken from points (a) and (b) in Figure\ \ref{fig:bright_snakes}, i.e., $\mu=-0.205$. The stable bright soliton was perturbed initially.}       	\label{fig:dynamics_bright}
	\end{figure}
	
	\begin{figure*}[tbhp] 
		\centering		
		\subfloat[$\mu=-0.195$]{\includegraphics[scale=0.55]{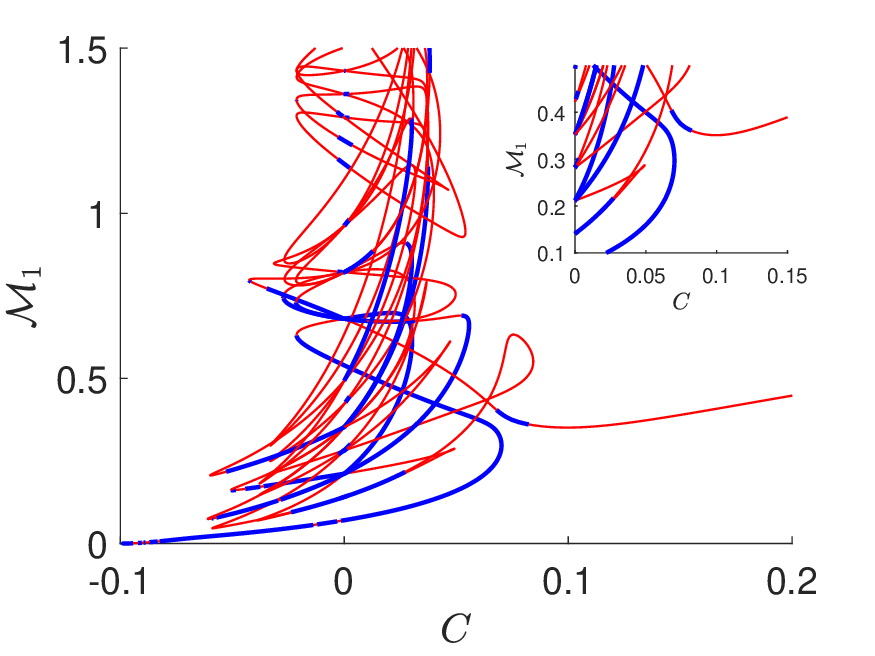}}
		\subfloat[$\mu=-0.194$]{\includegraphics[scale=0.55]{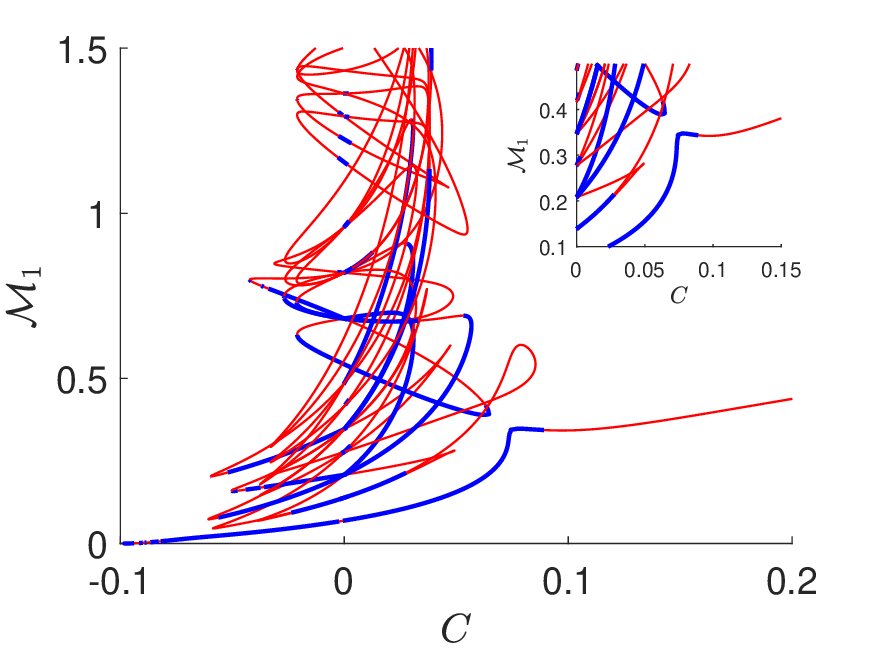}\label{fig:bright_2}}
		\caption{\textcolor{black}{
				The relationship of the solitons in the discrete and continuum limit for onsite solitons at \(\mu=-0.195\) and \(-0.194\). We plot the norm \(\mathcal{M}_1\) for varying \(C\). The insets show that there is branch switching, resulting in whether the fundamental solution can be continued rather 'smoothly' or not. In this diagram, stable solutions are represented by thick blue lines and unstable solutions are represented by thin red lines, enhancing visual clarity and aiding in understanding the stability dynamics of the solitons as \(C\) varies.}
		}        		\label{fig:bright_snakes_det}
	\end{figure*}
	
	\textcolor{black}{
		Figure \ref{fig:dynamics_bright_a} plots the typical time dynamics of the unstable bright soliton, initially profiled at point (a) in Figure \ref{subfig:bright_snake_c_0_10}. The soliton is expected to remain stationary (i.e., constant in time). However, due to its instability, it cannot maintain its structure. Over time, the instability manifests as an oscillation between neighboring sites for a prolonged period. We can observe the continuous release of wave radiation due to resonances with the continuous spectrum, possibly leading to the soliton's eventual self-destruction. In contrast, stable solitons under similar settings would maintain their shape over time without the oscillatory behavior or radiation release. A typical dynamics is shown in Figure \ref{fig:dynamics_bright_b}, which is initially taken from the solution at point (b) in Figure \ref{subfig:bright_snake_c_0_10}, with a small random perturbation. Such stable solutions exhibit persistent structural integrity due to the absence of resonances with the continuous spectrum, thus effectively preserving their initial form without significant deformation or decay. We observe radiations because of the initial perturbation.		 }
	
	The fundamental onsite and offsite bright solitons are simple in the uncoupled limit. We are also interested in whether these solutions can be continued to the continuum limit. For that, we fix the parameter $\mu$ and vary $C$ instead. We plot the norm $\mathcal{M}_1$ of the onsite soliton as a function of the coupling strength $C$ in Figure\ \ref{fig:bright_snakes_det} for two values of parameter $\mu$, i.e., $\mu=-0.195$ and $-0.194$. \textcolor{black}{We provide the bifurcation diagram for a restricted range of the coupling constant $C$ to ensure clarity in the presentation. Despite this limitation, we have continued the bifurcation curve up to significantly large values of $C$, far exceeding unity, and have not encountered any additional complex existence curves beyond those displayed.}
	
	For the latter value of $\mu$, solutions at the uncoupled and the continuum limit are connected straightforwardly. A slight change in the parameter $\mu$ can make a branch switching that yields a complicated bifurcation diagram. It is shown in Figure\ \ref{fig:bright_2}, where the connection between the continuum and uncoupled limits is no longer straightforward. 
	\textcolor{black}{We have identified critical values of the parameter \(\mu\), specifically \(\mu = -0.195\) and \(\mu = -0.194\), which separate two distinct behavioral cases in our analysis. These critical values delineate thresholds at which the system's properties undergo significant transitions. We have clarified that these thresholds are not directly related to any of the previously mentioned special values \({\mu^{(j)}}\), where \(j=0, 1, M\).}

	\textcolor{black}{The complex bifurcation diagrams reveal intricate behaviors of quantum droplets under various conditions. For instance, altering the chemical potential $\mu$ shifts the balance between dispersion and nonlinearity, resulting in different localized states. These insights are crucial for applications in areas such as Bose-Einstein condensates (BECs) and other contexts where discrete nonlinear Schrödinger (NLS) equations are applicable. In BECs, the coupling strength $C$ can be modified by adjusting the intensity and configuration of optical lattices created by counter-propagating laser beams or tuning the interaction strength between atoms using Feshbach resonances. This tuning affects the scattering length and, consequently, the coupling constant \cite{Porter2009}.}
	
	\subsection{Dark solitons}
	
	Next, we consider dark solitons that bifurcate from the turning point (saddle-node bifurcation) between the ``lower'' ${u^{(1)}}$ and ``upper'' uniform solutions $u^{(2)}$ at point ${\mu^{(1)}}$. They also have two fundamental localized solutions, i.e.,\ onsite and offsite states. They are formed by two bistable states from the uniform solutions, i.e., the zero solution ${u^{(0)}}$ as the ``dip'' state and the solution ${u^{(2)}}$ as the ``background'' state, i.e., the opposite of the bright solitons case.
	
\begin{figure*}[tbhp!] 
		\centering
		\subfloat[$C=0.1$]{\includegraphics[scale=0.55]{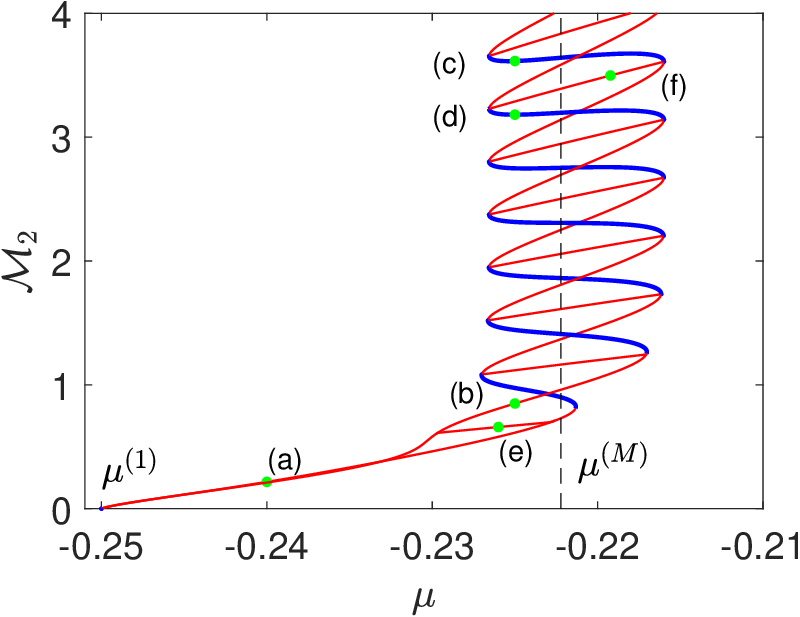}}\,
		\subfloat[$C=1$]{\includegraphics[scale=0.55]{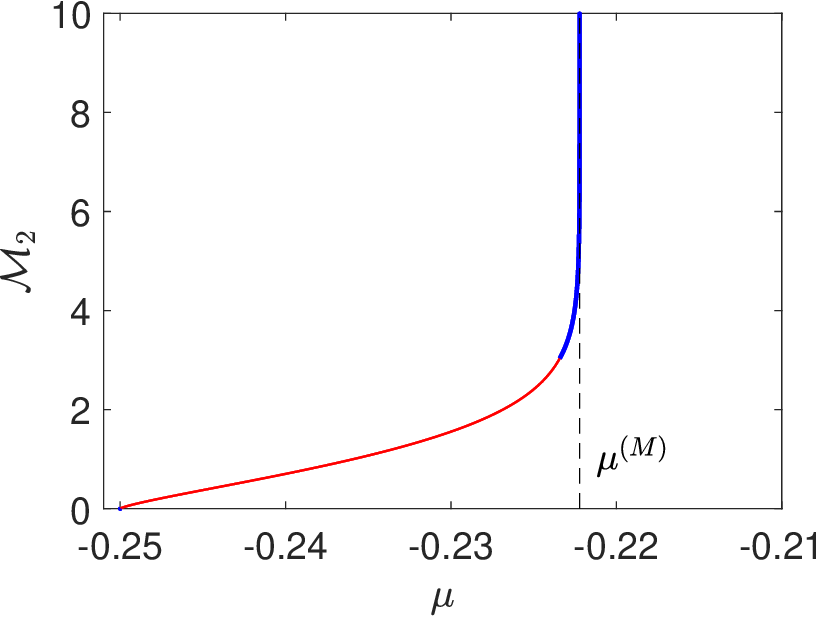}}
		\caption{Bifurcation diagram of dark solitons of Equation\ \eqref{eq:dnls23} for $C=0.1$ and $C = 1$. 
			We plot the norm $\mathcal{M}_2$ for varying $\mu$. There are two pairs of snaking branches for onsite and offsite solutions, {which are connected by ladders}. 
			\textcolor{black}{The thick blue lines represent stable, while the thin red lines depict unstable solutions. The corresponding solutions at points shown in panel (a) are plotted in Figure~\ref{fig:dark_snakes_prof}. Additionally, the critical value \( \mu^{(M)} \) represents the Maxwell point, where the energies of competing stable states are equal. 
		}} 		\label{fig:dark_snakes}
	\end{figure*}

\begin{figure*}[tbhp!] 
		\centering
		\subfloat[]{\includegraphics[scale=0.55]{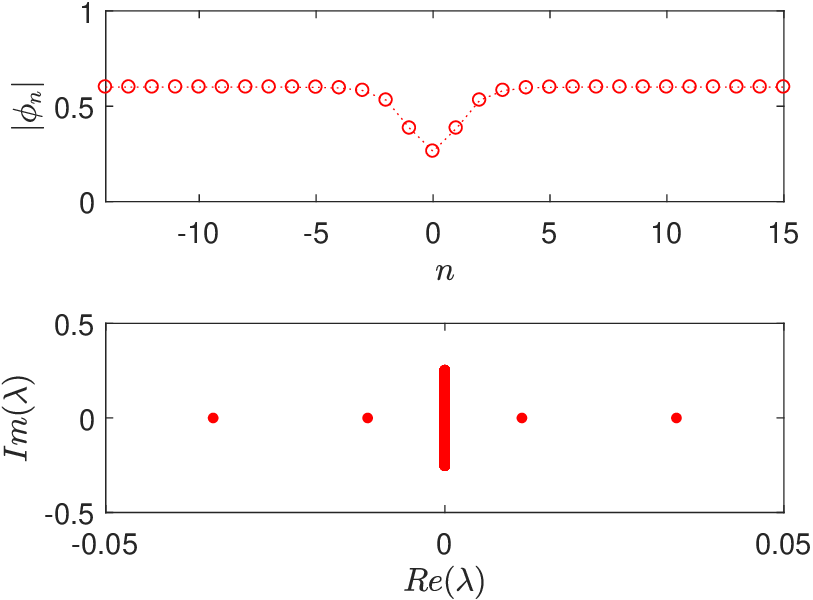}\label{subfig:prof_dark_a}}\,
		\subfloat[]{\includegraphics[scale=0.55]{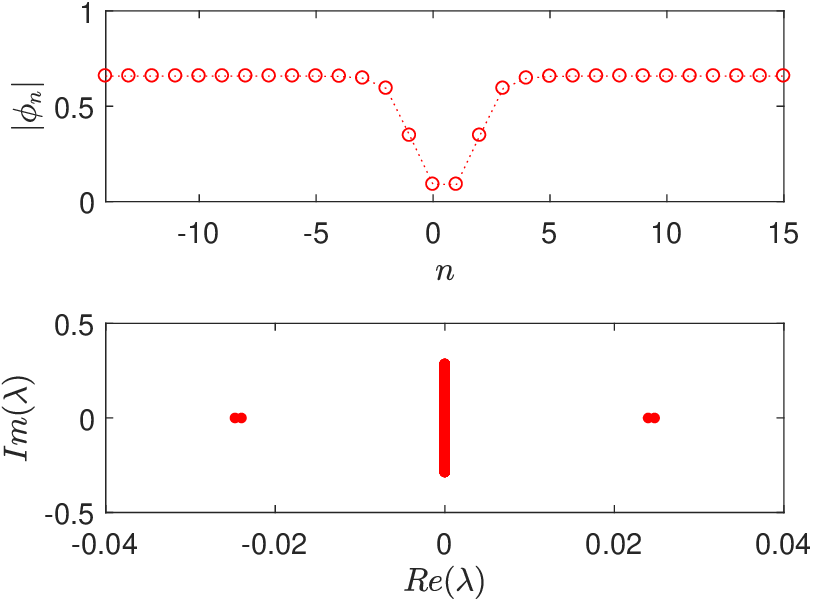}}\\
		\subfloat[]{\includegraphics[scale=0.55]{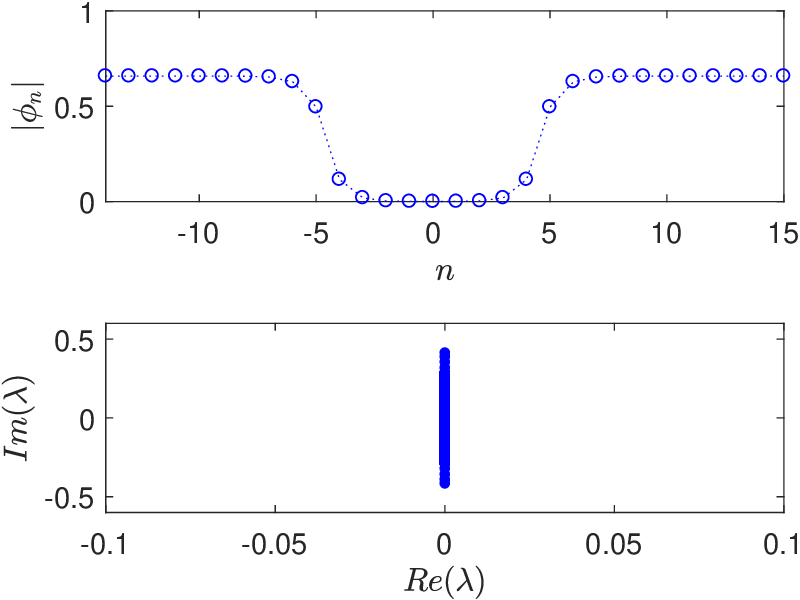}}\,
		\subfloat[]{\includegraphics[scale=0.55]{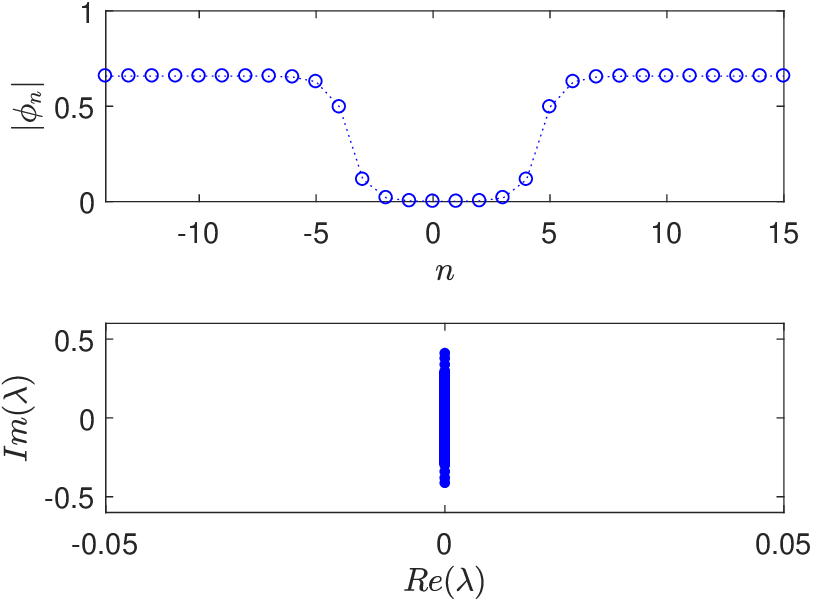}}\\
		\subfloat[]{\includegraphics[scale=0.55]{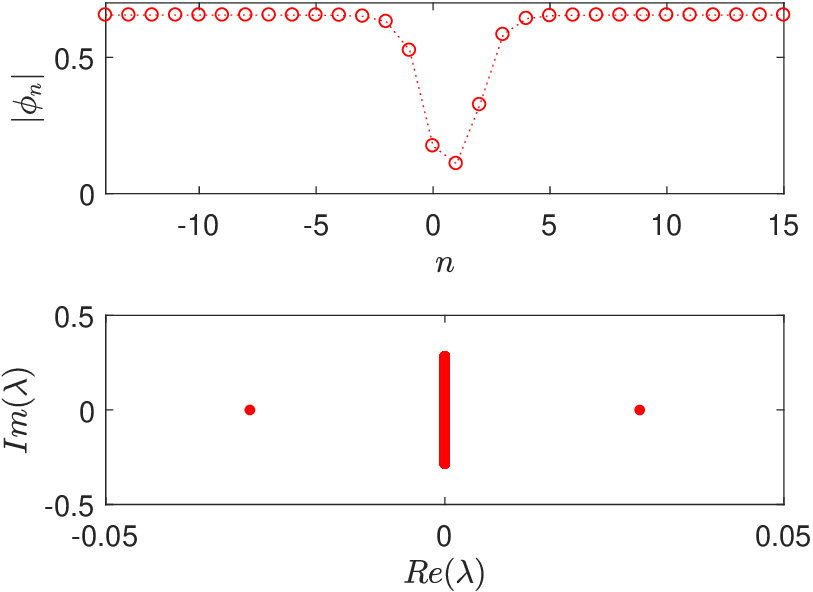}}\,
		\subfloat[]{\includegraphics[scale=0.55]{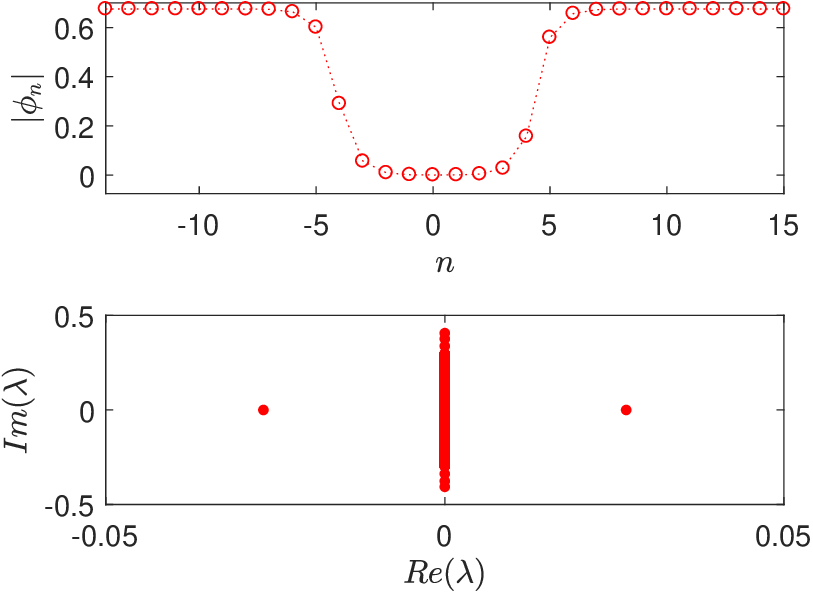}}
		\caption{Plots of solution profiles along the bifurcation diagrams in Figure\ \ref{fig:dark_snakes} and their spectra in the complex plane. {Panels [(a), (c)],  [(b), (d)],  and [(e), (f)] indicate onsite, offsite, and ladder solutions, respectively}.  \textcolor{black}{Blue and red colors represent stable and unstable solutions, respectively, consistent with the color scheme used in Figure \ref{fig:dark_snakes}.}} \label{fig:dark_snakes_prof}
	\end{figure*}

	Using the same numerical continuation for varying $\mu$ as in the previous section, we obtain bifurcation diagrams of the localized solutions that show a snaking structure similar to those for bright solitons, see Figure\ \ref{fig:dark_snakes}. Here, we use a 'modified' norm or ``mass'', where first we shift the solution background to zero \cite{yulin2010discrete,yulin2011snake}, i.e.,
	\begin{equation}
		\mathcal{M}_2=\sum_{n = -\infty}^{\infty} = \left|\phi_n-\phi_\infty\right|^2, \qquad \text{where} \quad \phi_{\infty} = \lim_{n\to\pm\infty} u_n.
	\end{equation}
	\textcolor{black}{We also plot in Figure \ref{fig:dark_snakes_prof} several solution profiles of onsite, offsite, and asymmetric localized solutions along the bifurcation diagrams, similar to Figure~\ref{fig:bright_snakes_prof} at various values of $\mu$, but for quantum bubbles.}
	
\begin{figure}[tbhp!] 
		\centering
		\subfloat[]{\includegraphics[scale=0.5]{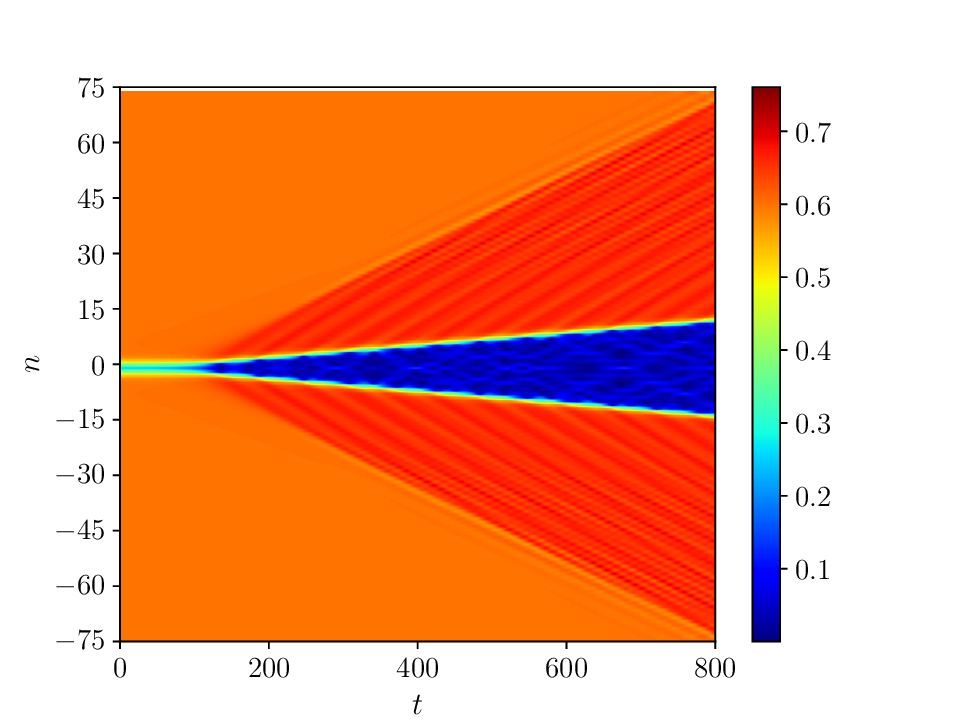}\label{fig:dynamics_dark_a}}\hspace{-1cm}
		\subfloat[]{\includegraphics[scale=0.5]{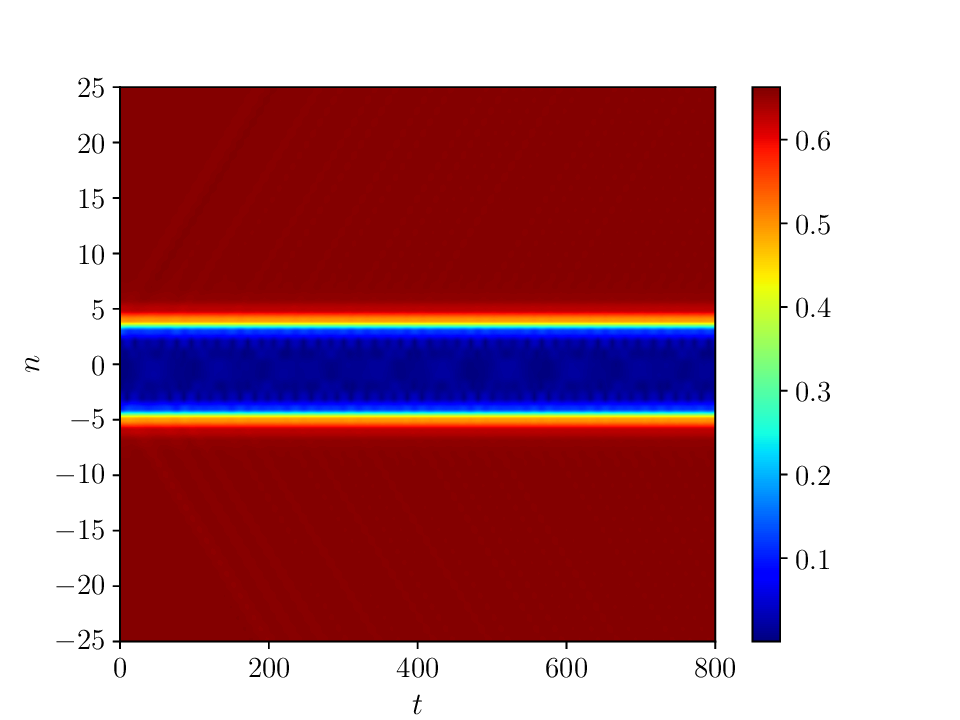}\label{fig:dynamics_dark_b}}
		\caption{Top view of time dynamics of (a) unstable and (b) stable dark solitons with the initial profile taken from points (a) and (d) in Figure\ \ref{fig:dark_snakes}, i.e., $\mu=-0.24$ and $\mu=-0.225$, respectively. The stable dark soliton was perturbed initially.} 		\label{fig:dynamics_dark}
	\end{figure}

	
	\textcolor{black}{
		Figure\ \ref{fig:dynamics_dark_a} shows the typical time evolution of unstable quantum bubbles, starting from the solution in Figure\ \ref{subfig:prof_dark_a}. The bubble splits into two kinks moving in opposite directions.  Radiation is emitted at a velocity larger than the kinks. The plateau of the trivial zero state becomes larger. This occurs because, at this value of \(\mu\), the energy of the zero state is lower than that of the nonzero state $u^{(2)}$. We also show in Figure\ \ref{fig:dynamics_dark_b} the typical dynamics of a stable dark soliton that is initially taken from point (d) in Figure\ \ref{subfig:prof_dark_a}. The scenario is markedly different for stable solutions, where they 
		maintain their structural integrity over time. They do not exhibit splitting, significant motion, or radiation leakage because their energy configurations stabilize their original forms. The slight radiation seen in Figure \ref{fig:dynamics_dark_b} is from the random perturbation we impose on the initial solution. 
	}
	
\begin{figure}[tbhp!] 
		\centering
		{\includegraphics[scale=0.55]{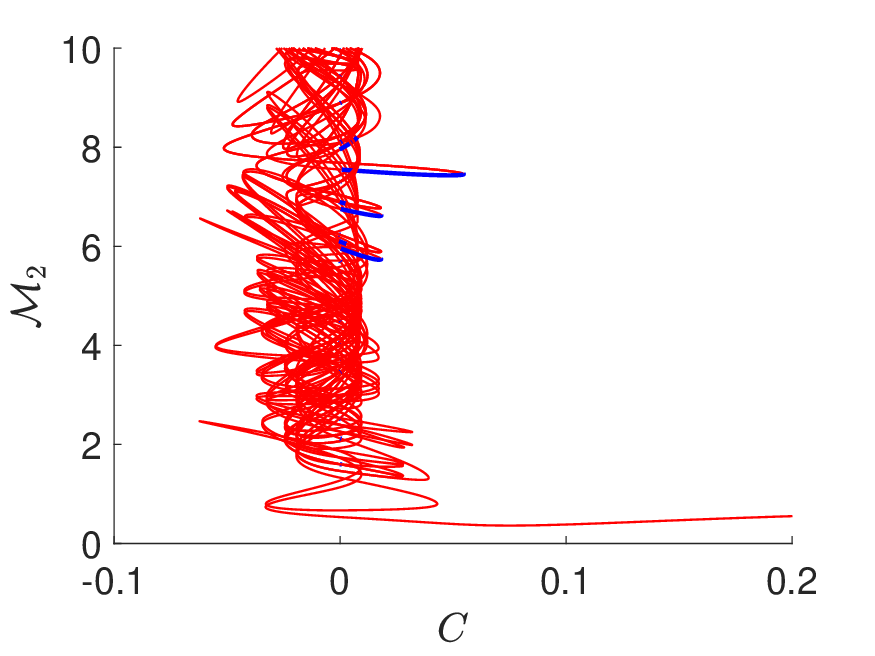}}
		\caption{{Similar to Fig.\ \ref{fig:bright_snakes_det}, this figure presents the bifurcation diagram of an onsite dark soliton at fixed $\mu=-0.233$ and varying coupling parameter $C$. The fundamental solution transitions smoothly from the discrete case to the continuum limit (i.e., the lowest branch).}}
		\label{fig:bifur_dark_mu_0_233}
	\end{figure}

	Similar to bright solitons, the fundamental onsite and offsite dark solitons are simple in the uncoupled limit. We investigate whether these solutions can be extended to the continuum limit by fixing the parameter $\mu$ and varying the coupling strength $C$. Figure \ref{fig:bifur_dark_mu_0_233} illustrates the norm $\mathcal{M}_2$ of the onsite soliton as a function of $C$ for $\mu=-0.233$. Offsite solitons also show similar diagrams (that we do not show here for simplicity). The result demonstrates a direct connection between solutions at the continuum and uncoupled limits, accompanied by an irregular snaking like that in Figure\ \ref{fig:bright_snakes_det} for bright solitons. 
	
	\section{The width of the pinning region}          \label{sec4}
	
	In this section, we will study the width of the snaking region for bright and dark solitons in Figures\ \ref{fig:bright_snakes} and \ref{fig:dark_snakes} as a function of the coupling constant $C$. However, their widths are identical for the same coupling strength $C$. We plot the width that is calculated numerically in Figure \ref{fig:pinning_width}. This is because bright and dark solitons become two well-separated kink solutions up in the snaking curves. The snaking boundaries are, therefore, determined by the existence of kink solutions. 
	In the following, we derive asymptotic approximations of the width of the snaking region. The analysis is divided into two regions, i.e., small and large couplings.

	\subsection{Small coupling case}
	
	When the coupling strength $C$ is weak, up to the leading order, the solutions effectively consist of only three states, i.e., uniform state ${u^{(2)}}$, zero state ${u^{(0)}}$, and an interface (active site) $\xi$ \cite{kusdiantara2017homoclinic,kusdiantara2019snakes,kusdiantara2022snakes}. Using this approximation, only three states will be involved in the dynamics for small coupling $C$. Thus, we can write Equation\ \eqref{eq:dnls23_ti} into a simple cubic equation in $\xi$:
	\begin{equation}
		f(\xi):=-\mu\xi-\frac{C}{2}\left(u^{(2)}-2\xi\right)+\xi^3-\xi^2=0. 	  \label{eq:active_site}
	\end{equation}
	
	According to the Fundamental Theorem of Algebra, Equation\ \eqref{eq:active_site} will have three roots in $\xi$. The roots relevant to our study are the positive ones. As $\mu$ varies, two roots will collide in a saddle-node bifurcation. This condition corresponds to the boundaries of the snaking region. The condition for the collision occurs when a local maximum or minimum of the function $f(\xi)$ crosses the horizontal axis. The critical points of $f(\xi)$ must satisfy the following equation
	\begin{equation}
		f^\prime(\xi):=-\mu+C+3\xi^2-2\xi=0,
	\end{equation}
	and hence the critical points are given by 
	\begin{equation}
		\xi=\frac{1}{3}\pm\frac{1}{3}\sqrt{1+3\mu-3C}.              \label{eq:active_site_point}
	\end{equation}
	Substituting Equation~\eqref{eq:active_site_point} into Equation\ \eqref{eq:active_site} and solving the resulting expression using the Newton-Raphson method, we plot the small $C$ approximation shown in Figure\ \ref{fig:pinning_width}, as the green dashed line. Good agreement is obtained for small coupling strength \( C \). The approximation diverges significantly from the numerical results for $C\gtrapprox0.1$. This limitation is caused by our approximation, which assumes only one interface. We can improve it straightforwardly by including more active sites.
	
	\subsection{Large coupling case}
	
	There are several published works on snaking in a discrete system when the coupling between sites is large~\cite{matthews2011variational,dean2015orientation,susanto2018snakes}. While Matthews and Susanto~\cite{matthews2011variational} and Susanto~et~al.~\cite{susanto2018snakes} used a variational approach to capture the snaking that yields the correct scaling of the width, Dean~et~al.~\cite{dean2015orientation} employed beyond-all-order asymptotics that gave both the right scaling and prefactor of the exponentially small phenomenon. Nevertheless, in the following, we will use the method in~\cite{susanto2018snakes} to obtain the width of the pinning region as it is particularly simple but accurate enough. 
	
	\begin{figure*}[tbhp!] 
		\centering
		\subfloat[Normal scale]{\includegraphics[scale=0.53]{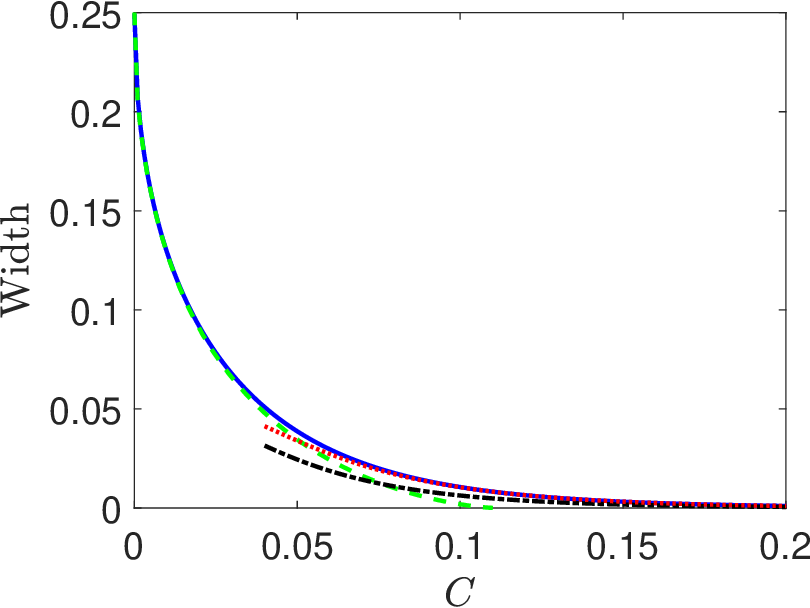}\label{subfig:pinning_region_v2}}\,
		\subfloat[Logaritmic scale]{\includegraphics[scale=0.53]{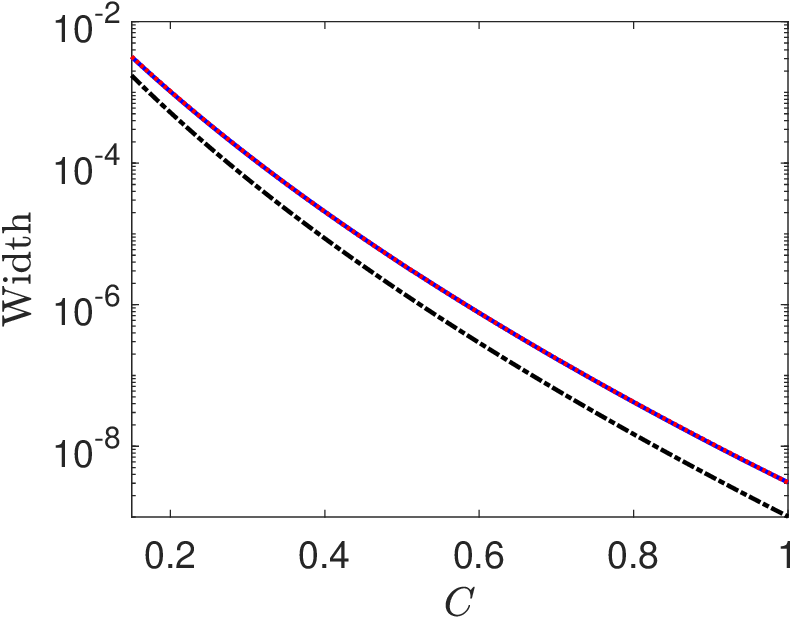}\label{subfig:pinning_region_log_v2}}
		\caption{(Color online). (a) and (b) illustrate the pinning width of the snaking region on normal and logarithmic scales, respectively, as a function of the coupling constant $C$ in different regions. The blue solid curve is a numerical result. The green dashed and the black dash-dotted lines represent the approximations given in Equation \eqref{eq:active_site} and 
			Equation \eqref{exp1}, respectively. We also plot the curve fit in Equation \eqref{exp3} as the red dotted line.} \label{fig:pinning_width}
	\end{figure*}
	
	Following the line of reasoning from~\cite{susanto2018snakes}, the approach starts with approximating the discrete system \eqref{eq:dnls23_ti} with the following continuous equation with a periodic potential 
	\begin{equation}
		\frac C2\tilde{A}_{xx}+\sum_{n=-\infty}^\infty\delta(x-n) \; F \left[\tilde{A}(x) \right] =0,		\label{gov4}
	\end{equation}
	where $A_n=\tilde{A}(x=n)$ and $F(\tilde{A})=\mu \tilde{A} +|\tilde{A}|\tilde{A}-\tilde{A}^3$. The proof that Equation~\eqref{gov4} is equivalent to Equation~\eqref{eq:dnls23_ti} in a 'weak' sense is as follows. 
	
	First, from \eqref{gov4}, we obtain that $\tilde{A}_{xx}(n<x<n+1)=0$. Upon integration, this is the same as $\tilde{A}_{x}=\text{constant}$ for $n<x<n+1$. Hence, we may write that at $x=n+1/2$,
	\begin{equation}
		\tilde{A}_x(n+1/2)=\tilde{A}(n+1)-\tilde{A}(n). 		\label{s4}
	\end{equation}
	Next, we integrate \eqref{gov4} from $x=n-1/2$ to $x=n+1/2$ to obtain
	\begin{equation}
		\frac C2\left[\tilde{A}_{x} \left(n+ \frac{1}{2} \right) - \tilde{A}_{x} \left(n- \frac{1}{2} \right)\right]+ F \left[\tilde{A}(n) \right]=0.       	\label{s5}
	\end{equation}
	Using Equation\ \eqref{s4}, Equation\ \eqref{s5} becomes the lattice equation~\eqref{eq:dnls23_ti}.
	
	The next approximation comes from the fact that the Fourier transform of the Dirac comb 
	\begin{equation*}
		\Sha(x)=\sum_{n=-\infty}^\infty\delta(x-n)    
	\end{equation*}
	is given by 
	\begin{equation*}
		\mathcal{F}\{\Sha(x)\}=1+2\sum_{k=1}^\infty\cos(2\pi kx),    
	\end{equation*} 
	which converges to the Dirac comb non-uniformly. Taking only the first harmonic, Equation~\eqref{gov4} then becomes
	\begin{equation}
		\frac C2\tilde{A}_{xx}+ \left[1+2\cos(2\pi x) \right] \left(\mu \tilde{A} +|\tilde{A}|\tilde{A}-\tilde{A}^3\right)=0,		\label{govt}
	\end{equation}
	which can be expected to approximate \eqref{eq:dnls23_ti} in the large coupling case for $C\gg1$.
	
	Without the periodic potential $2\cos(2\pi x)$, there is a kink solution $\Phi_K$ given by~\cite{katsimiga2023interactions,katsimiga2023solitary}
	\begin{equation}    
		\Phi_K=\frac13\left[1+\tanh \left(\frac{x}{3\sqrt{C}}\right) \right]    		\label{kink}
	\end{equation}
	at the Maxwell point $\mu^{(M)}=-2/9$. 
	
	Following the argument in~\cite{susanto2011variational,matthews2011variational}, we will approximate the symmetric solutions along the snaking structure by
	\begin{equation}
		\tilde{A}(x)=\frac13\left[1+\tanh \left(\frac{\pm\left(|x-\phi|-L\right)}{3\sqrt{C}}\right) \right], 		\label{fr}
	\end{equation}
	where $\phi$ is the phase-shift distinguishing the two branches, i.e., $\phi = 0$ and $\phi = 1/2$ for the on-site and intersite solutions, respectively. The ``$\pm$'' sign corresponds to the quantum bubbles and droplets. The constant $L$ in~\eqref{fr} denotes the plateau length, an unknown variable. We then require \eqref{fr} to be an optimal solution of Equation \eqref{govt}. Using the standard variational argument, this implies that $L$, as the only unknown parameter, must satisfy the following equation (see, e.g., \cite{dawes2013variational})
	\begin{equation}
		\int_{-\infty}^\infty \left[\frac C2\tilde{A}_{xx}+(1+2\cos(2\pi x))F(\tilde{A})\right] \frac{\partial \tilde{A}}{\partial L}\,dx=0,        		\label{solvb}
	\end{equation}
	where $\mu$ is set to be near the Maxwell point $\mu^*$, i.e.,\ $\mu=-2/9+\Delta\mu$. Equation \eqref{solvb} can be simplified at the leading order for $L\gg1$ to
	\begin{align}
		\displaystyle
		\Delta\mu&=\lim_{L\to\infty} \frac{\displaystyle C\int_{-\infty}^\infty \cos(2\pi x)\tilde{A}_{xx}\frac{\partial \tilde{A}}{\partial L}\,dx}{\displaystyle \int_{-\infty}^\infty (1+2\cos(2\pi x)) \tilde{A}\frac{\partial \tilde{A}}{\partial L}\,dx}\approx
		\lim_{L\to\infty} \frac{\displaystyle C\int_{-\infty}^\infty \cos(2\pi x)\tilde{A}_{xx}\frac{\partial \tilde{A}}{\partial L}\,dx}{\displaystyle \int_{-\infty}^\infty \tilde{A}\frac{\partial \tilde{A}}{\partial L}\,dx}.  		\label{solvb1}
	\end{align}
	
	The integral in the numerator \eqref{solvb1} with the cosine factor can be evaluated using tables of integrals (e.g.,~\cite{gradshteyn1988tables}) or symbolic computation software, such as Mathematica, that will involve the hypergeometric function. Nevertheless, computations can also be done using the residue integration method (see, e.g.,~\cite{kreyszig2007advanced}), i.e.\
	\begin{equation}
		\int_{-\infty}^{\infty} f(x)\cos(2\pi x)\,dx=-2\pi\sum \text{Im}\,\text{Res}\left[f(x)e^{i2\pi x}\right],
	\end{equation}
	where the summation is over all the infinitely many poles of the function that lie in the upper half of the complex plane. It is easy to note that the poles are given by 
	\begin{equation*}
		x_k=\phi+\frac{3i\pi}2\sqrt{C}\left(1+2k\right),\quad k\in\mathbb{Z}^{\geq0}.
	\end{equation*}
	
	However, in the following, we will only consider the principal one\added{, i.e., } $k=0$, which should already give an accurate approximation to the integral \eqref{solvb1} for a large value of $C$. We, therefore, obtain that 
	\begin{align*}
		\lim_{L\to\infty}\int_{-\infty}^\infty \cos(2\pi x)\tilde{A}_{xx}\frac{\partial \tilde{A}}{\partial L}\,dx &=
		{\mp\frac8{27}{e^{-3{\pi }^{2}\sqrt{C}}}{\pi }^{3} \left( 9{C}{\pi }^{2}+1 \right) }\sin(2\pi L),\\
		\lim_{L\to\infty}\int_{-\infty}^\infty \tilde{A}\frac{\partial \tilde{A}}{\partial L}\,dx &= \mp \frac{2}{9}.
	\end{align*}
	The width $W$ of the snaking region is then given by
	\begin{align}
		W &= 2\Delta\mu = {\frac{8}{3}{\pi }^{3}{C}{e^{-3{\pi }^{2}\sqrt{C}}} \left( 9{C}{\pi }^{2}+1 \right)}
		\sim  24{\pi }^{5}{C}^{2}{e^{-3{\pi }^{2}\sqrt{C}}},\label{exp1}
	\end{align}
	which is exponentially small.
	
	\textcolor{black}{
		We show in Figure\ \ref{fig:pinning_width} the width of the snaking region computed numerically and our approximation \eqref{exp1}. Note that Figure~\ref{subfig:pinning_region_v2} does not allow one to check the accuracy of \eqref{exp1} because of the fast exponential decay as $C$ increases. In Figure~\ref{subfig:pinning_region_log_v2}, we depict the comparison in a logarithmic plot, where one can see good agreement between them. However, we also note that the approximation slowly deviates from the numerical computations as $C$ increases. 
	} Using the function
	\begin{align}
		W =  \alpha C^\beta e^{-{3\pi^2\sqrt{C}}},\label{exp3}
	\end{align}
	we curve-fit the numerical result where we obtain $\alpha=22,212.24$ and $\beta=2.26$. While $\beta$ is relatively close to the analytical result, with around 13\% difference, the value of the factor $\alpha$ is three times larger than the analytical approximation, i.e., $\alpha > 72\pi^5 \approx 22,033.42$. The difference may be attributed to taking the first harmonic only in Equation~\eqref{govt}. 
	
	\section{Conclusions}           \label{sec5}
	
	We have considered the existence, stability, and dynamics of discrete quantum droplets and bubbles in 
	the discrete NLS equation where the nonlinearity consists of quadratic and cubic terms. In particular, we discussed the mechanism of the homoclinic snaking phenomenon extensively and calculated the associated pinning region width. 
	
	For small coupling strength, the system can be reduced into three states only: non-zero plateau, zero background, and one interface state. Our analysis suggests that the width of the pinning region associated with the homoclinic snaking is inversely proportional to the strength of the coupling constant between the sites. Analytical and numerical results show remarkable agreements.
	
	When the coupling constant between the sites is sufficiently strong, we employed a variational method to investigate the homoclinic snaking phenomenon. We obtained that the width is quadratically proportional to the coupling constant and depends on an exponentially small term where its power is proportional to the square root of the coupling strength. The numerical simulation further confirms that our analytical approximation for the power of the coupling constant was undershot by around 13\%. 
	
	\textcolor{black}{For future work, it is interesting to analyze the snaking properly using beyond-all-order asymptotics. An asymptotic analysis of the critical eigenvalues along the snaking is also intriguing and remains to be done. Another interesting problem is semidiscrete quantum droplets studied recently by Zhang et al.\ \cite{zhang2019semidiscrete}, where the system is continuous in one direction but discrete in the other, representing transverse hopping between nearly one-dimensional traps. The system was shown to support multistable states within a finite interval of parameters. That may indicate that the states are connected in a snaking bifurcation diagram. It is important to analyze this prediction further and evaluate if our findings reported herein are applicable, which may yield new theoretical frameworks.}
	
	\section*{Acknowledgement}
	\noindent This work was supported by a Faculty Start-Up Grant (No.\ 8474000351/FSU-2021-011) by Khalifa University. HS also acknowledged support from Khalifa University through a Competitive Internal Research Awards Grant (No.\ 8474000413/CIRA-2021-065).	RK also acknowledges Riset ITB 2023 (7215/IT1.B07.1/TA.00/2022). 
	
	\section*{Conflict of interests}
	\noindent The authors declare that they have no conflict of interests.
	
	\bibliographystyle{elsarticle-num}

\end{document}